\documentclass[twocolumn, final]{autart}    

\usepackage[colorlinks = true,
            linkcolor = blue,
            urlcolor  = blue,
            citecolor = blue,
            anchorcolor = blue]{hyperref}
\usepackage{graphicx}
\usepackage{subfig}
\usepackage{amsmath}
\usepackage{amssymb}
\usepackage{amsfonts}
\usepackage{enumerate}
\usepackage{mathtools}
\usepackage{colortbl}
\usepackage{color}
\usepackage{subfig}
\usepackage{nicefrac}
\usepackage{stfloats}
\usepackage{booktabs}
\usepackage{colortbl}
\usepackage{siunitx}
\usepackage[title]{appendix}

\usepackage{listings}
\definecolor{green_new}{RGB}{0, 126, 0}

\begin{document}

\begin{frontmatter}

\title{On the smoothness  of nonlinear system identification}

\author[grad_program,uu]{Ant\^{o}nio H. Ribeiro}\ead{antonio-ribeiro@ufmg.br},
\author[uu,tue]{Koen Tiels},
\author[uu]{Jack Umenberger},
\author[uu]{Thomas B. Sch\"on},
\author[elet_dep]{Luis A. Aguirre}

\address[grad_program]{Graduate  Program in  Electrical  Engineering,
   Universidade Federal de Minas Gerais, Brazil}  
\address[elet_dep]{Dept. of Electronic Engineering,
   Universidade Federal de Minas Gerais, Brazil}
\address[uu]{Dept. of Information Technology, Uppsala University, Sweden}  
\address[tue]{Dept. of Mechanical Engineering, Eindhoven University of Technology, The Netherlands}


\vspace{-0.5cm}
\begin{abstract}                          

We shed new light on the \textit{smoothness} of optimization problems arising in prediction error parameter  estimation of linear and nonlinear systems. We show that for regions of the parameter space where the model is not contractive, the Lipschitz constant and $\beta$-smoothness of the objective function might blow up exponentially with the simulation length, making it hard to numerically find minima within those regions or, even, to escape from them. In addition to providing theoretical understanding of this problem, this paper also proposes the use of multiple shooting as a viable solution.  The proposed method minimizes the error between a prediction model and the observed values. Rather than running the prediction model over the entire dataset, multiple shooting splits the data into smaller subsets and runs the prediction model over each subset, making the simulation length a design parameter and making it possible to solve problems that would be infeasible using a standard approach. The equivalence to the original problem is obtained by including constraints in the optimization. The new method is illustrated by estimating the parameters of nonlinear systems with chaotic or unstable behavior, as well as neural networks. We also present a comparative analysis of the proposed method with multi-step-ahead prediction error minimization.
\end{abstract}

\end{frontmatter}

\begin{figure}[b]
\noindent\textbf{\footnotesize Please cite:}\textit{\footnotesize A. H. Ribeiro, K. Tiels, J. Umenberger, T. B. Schön, and L. A. Aguirre, “On the smoothness of nonlinear system identification,” Automatica, vol. 121, 109158, Nov. 2020, doi: 10.1016/j.automatica.2020.109158.}

\begin{lstlisting}
@article{ribeiro_smoothness_2020,
title = {On the smoothness  of nonlinear system identification},
author = {Ribeiro, Ant{\^o}nio H. and Tiels, Koen and Umenberger, Jack and Sch{\"o}n, Thomas B. and Aguirre, Luis A.},
year = {2020},
volume = {121},
pages = {109158},
journal = {Automatica}
doi = {10.1016/j.automatica.2020.109158},
}
\end{lstlisting}
\end{figure}

\section{Introduction}

Prediction error methods~\cite{ljung_system_1998} are a widespread class of methods for parameter estimation of dynamic models, which estimate the parameters by minimizing the error between predicted and measured trajectories. Many well-known estimation methods fit into this framework, such as minimizing the one-step-ahead prediction error or the free-run-simulation error. While the classical literature focuses primarily on the estimation of linear systems~\cite{ljung_system_1998}, the framework is general and enjoys appealing asymptotic properties for the general nonlinear setup~\cite{ljung_convergence_1978}. 

Minimizing the one-step-ahead prediction usually yields an easier optimization problem, the minimization of the free-run simulation error  or other recurrent structures, however, may produce more accurate models. These recurrent models often have a smaller generalization error~\cite{ribeiro_parallel_2018},~\cite{aguirre_prediction_2010} and better capability when it comes to long-term  prediction~\cite{su_longterm_1992}. Minimizing recurrent structures is used, for instance, to improve the model structure selection of polynomial models~\cite{piroddi_simulation_2008}, fine-tune parameters of nonlinear state-space~\cite{paduart_identification_2010} and block-oriented models~\cite{schoukens_identification_2017}.

It is common knowledge among practitioners that the optimization problem resulting from a recurrent model structure is harder to solve~\cite{piroddi_identification_2003}. 
For linearly parametrized models and convex loss functions, minimization of one-step-ahead prediction error leads to a convex optimization problem; for recurrent model structures, the ensuing optimization is, in general, non-convex, complicating the search for global optima.
Even during local optimization, recurrent model structures can lead to cost functions with poor smoothness properties, including many `jagged' local minima, cf. Fig.~\ref{fig:ms_logistic}(a) for an illustration. The current understanding of the relationship between model internal dynamics and the smoothness properties of the cost function is, however, imprecise and provides little insight into ways of circumventing the problem. Furthermore, the few studies that do investigate the objective function properties in this context are focused on linear systems, see e.g.~\cite{eckhard_cost_2017}.

The purpose of this paper is twofold. 
First, we aim to provide insight into the properties of the objective function arising in prediction error estimation problems in a general nonlinear setup.
Specifically, we show how the smoothness of the objective function depends on two factors: the simulation length and the decay rate of the recurrent part of the prediction model. Second, we illustrate how this theoretical insight might be leveraged for the design and analysis of practical system identification methods.

The use of \textit{multiple shooting} is analyzed in the context of  prediction error minimization. This technique reformulates the optimization problem that arises from  minimizing the difference between the output of a prediction model and the observed values. Rather than running the prediction model over the entire dataset, the multiple shooting formulation splits the dataset into smaller subsets and runs the prediction model over each subset. The equivalence with the original problem is obtained by including equality constraints in the optimization problem. This method results in a smoother objective function since it works with shorter simulations and prevents trajectories from diverging too much.

The multiple shooting formulation has reportedly provided improvements in the parameter estimation of ordinary differential  equations~\cite{bock_recent_1983},~\cite{baake_fitting_1992},~\cite{sarode_embedded_2015} and in the solution of optimal control~\cite{bock_multiple_1984},~\cite{carraro_indirect_2014}, \cite{geisert_trajectory_2016}. In the context of system identification, multiple shooting has been used for estimating polynomial nonlinear space-state models~\cite{vanmulders_two_2010} and output error models~\cite{ribeiro_shooting_2017} in settings where conventional methods fail to provide good solutions. Here, we extend this method to the entire class of prediction error methods. In addition, theoretical arguments are put forward to help understand why and when the proposed method is useful.
We also present a comparative analysis with multi-step-ahead prediction error minimization~\cite{farina_simulation_2011},~\cite{terzi_learning_2018}~\cite{farina_identification_2012}, showing strengths and weaknesses of each method at a conceptual level and, also, with numerical examples.

\section{Prediction error methods}
\label{sec:pred-error-meth}

While prediction error methods are widely known, they are often introduced from a linear perspective~\cite{ljung_system_1998}. In the present section, this is accomplished in a nonlinear setting.  

Consider the dataset $\mathcal{Z}^N = \{(\mathbf{u}[k],\mathbf{y}[k]),  k=1,2,\cdots,N\}$ containing $N$ measured inputs and outputs of a dynamical system.
Prediction error methods assume an internal model that delivers  output predictions $\hat{\mathbf{y}}[k]$, $k=1,2,\cdots,N$. A cost function is defined as the distance between predictions and measured values:
\begin{equation}
  \label{eq:cost_function}
  V = \frac{1}{N}\sum_{k=1}^N \|\mathbf{y}[k] - \hat{\mathbf{y}}[k]\|^2.
\end{equation}
The prediction model is, usually, parametrized by a parameter vector $\boldsymbol{\theta}$ and, although such a dependence is not made explicit in the notation, $\hat{\mathbf{y}}[k]$ depends upon $\boldsymbol{\theta}$. An estimate $\hat{\boldsymbol{\theta}}$ of the parameter vector may be obtained by minimizing $V$.

Next we assume a dynamical, stochastic, discrete-time system as the data generating process. Let:
\begin{eqnarray*}
    \underline{\mathbf{u}}[k] &=& (\mathbf{u}[k], \cdots, \mathbf{u}[k-n_u]), \\
    \underline{\mathbf{y}}[k-1] &=& (\mathbf{y}[k-1],  \cdots, \mathbf{y}[k-n_y]),
\end{eqnarray*}
where  $n_y$, $n_u$ are the maximum input and output lags. The examples we give next will differ in how they account for the noise in the model. But, in a noiseless situation, the data generating process for all of them corresponds to a difference equation $\mathbf{y}[k] = \mathbf{f}^*(\underline{\mathbf{y}}[k-1], \underline{\mathbf{u}}[k])$. The parametrized function $\mathbf{f}_{\boldsymbol{\theta}}$ is defined and, if the noise assumption and model structure are correct, the minimization of the cost function $V$ would yield the estimate $\hat{\boldsymbol{\theta}}$ such that, as $N \rightarrow \infty$, then $\mathbf{f}_{\mathbf{\hat{\boldsymbol{\theta}}}} \rightarrow \mathbf{f}^*$ (or to a function with different structure but the same performance in predicting the training dataset). See Appendix~\ref{sec:asymptotic-properties} for a more precise description of the asymptotic properties of nonlinear prediction error methods.

Choosing the ``true'' model structure (one for which there exists a $\boldsymbol{\theta}^*$  such that $\mathbf{f}_{\boldsymbol{\theta}^*} = \mathbf{f}^*$) might be impossible in a practical application. Nevertheless the assumption is not so restrictive as it might appear, since there exist families of universal approximator functions (e.g. neural networks and polynomials) for which the distance ${\|\mathbf{f}_{\boldsymbol{\theta}} - \mathbf{f}^*\|}$ might be made arbitrarily small within a compact set.

\subsection{Nonlinear ARX models}
\label{sec:narx}

The nonlinear ARX (\textit{autoregressive with exogenous input}) model encodes an output that is corrupted by white \textit{process} noise. That is, it assumes the data were generated by the stochastic discrete-time system:
\begin{equation}
    \mathbf{y}[k] = \mathbf{f}^*(\underline{\mathbf{y}}[k-1], \underline{\mathbf{u}}[k]) + \mathbf{v}[k],
\end{equation}
where $\mathbf{v}[k]$ is a zero-mean white noise. This assumption yields (cf. Appendix~\ref{sec:asymptotic-properties}) the prediction model:
\begin{equation}
    \label{eq:narx_optimal}
    \hat{\mathbf{y}}[k] = \mathbf{f}_{\mathbf{\boldsymbol{\theta}}}(\underline{\mathbf{y}}[k-1], \underline{\mathbf{u}}[k]).
\end{equation}
The minimization of the cost function in Eq.~(\ref{eq:cost_function}) for this prediction model yields an estimator with the desired asymptotic properties.

\subsection{Nonlinear OE models}
\label{sec:noe}

The OE (\textit{output error}) model encodes an output that is corrupted by white \textit{measurement} noise. That is, it is assumed that the data were generated by:
\begin{align*}
\bar{\mathbf{y}}[k] &= \mathbf{f}^*(\bar{\mathbf{y}}[k-1], \hdots, \bar{\mathbf{y}}[k-n_y],
      \underline{\mathbf{u}}[k]),\\
\mathbf{y}[k] &= \bar{\mathbf{y}}[k] + \mathbf{v}[k],
\end{align*}
where, again, $\mathbf{v}[k]$ is zero-mean white noise and $\bar{\mathbf{y}}[k]$ represents the noiseless output. This assumption yields the following prediction model:
\begin{align}
\tilde{\mathbf{y}}[k] &= \mathbf{f}_{\boldsymbol{\theta}}(\tilde{\mathbf{y}}[k-1], \hdots, \tilde{\mathbf{y}}[k-n_y],
      \underline{\mathbf{u}}[k]),\nonumber\\
\hat{\mathbf{y}}[k] &= \tilde{\mathbf{y}}[k].
    \label{eq:noe_optimal}
\end{align}
\noindent
Here $\tilde{\mathbf{y}}[k]$ represents an estimate of the noiseless output $\bar{\mathbf{y}}[k]$, which should approach the true value as ${\boldsymbol{\theta} \rightarrow \boldsymbol{\theta}^*}$.

\subsection{Nonlinear ARMAX models}
\label{sec:narmax}

The nonlinear ARMAX (\textit{autoregressive moving average with exogenous input}) model encodes an output that is corrupted by additive zero-mean \textit{process} noise. In this case, the propagation equation allows the noise term $\mathbf{v}[k]$ to be propagated by the dynamics:
\begin{equation*}
    \mathbf{y}[k] = \mathbf{f}^*(\underline{\mathbf{y}}[k-1], \underline{\mathbf{u}}[k], \mathbf{v}[k-1], \hdots,
\mathbf{v}[k-n_v]) + \mathbf{v}[k].
\end{equation*}
This assumption allows the model to account for some forms of colored process noise and results in the following prediction model:
\begin{align}
&\tilde{\mathbf{v}}[k] = \mathbf{y}[k] - \mathbf{f}_{\boldsymbol{\theta}}(\underline{\mathbf{y}}[k-1],
\underline{\mathbf{u}}[k],
\tilde{\mathbf{v}}[k-1], \hdots,
\tilde{\mathbf{v}}[k-n_v]) \nonumber\\
 &\hat{\mathbf{y}}[k] = \mathbf{f}_{\boldsymbol{\theta}}(\underline{\mathbf{y}}[k-1],
                  \underline{\mathbf{u}}[k],
                  \tilde{\mathbf{v}}[k-1], \hdots,
                  \tilde{\mathbf{v}}[k-n_v]). \label{eq:narmax_optimal}
\end{align}
\noindent Here $\tilde{\mathbf{v}}[k]$ represents an estimate of the noise corrupting the system and will approach the true noise if the estimated parameter vector approaches the true parameter value $\boldsymbol{\theta}^*$.

\subsection{General nonlinear state-space framework}

From now on, we will focus on a  state-space representation that is general enough to encompass the prediction model from the above examples (for an appropriate choice of the functions $\mathbf{h}$ and $\mathbf{g}$). For this representation, the predicted output is given by:
\begin{subequations}
  \label{eq:nlss}
  \begin{align}
  \mathbf{x}[k] &= \mathbf{h}(\mathbf{x}[k-1], \underline{\mathbf{z}}[k];
                    \boldsymbol{\theta}),\label{eq:nlss-state-equation}\\
  \hat{\mathbf{y}}[k] &= \mathbf{g}(\mathbf{x}[k],  \underline{\mathbf{z}}[k]; \boldsymbol{\theta}),\label{eq:nlss-output-equation}
  \end{align}
\end{subequations}
where $\textbf{x}[k]$ denotes the internal state vector at instant $k$. For the ARX model the transition state would be empty  $\mathbf{x} = \varnothing$; for the OE model  ${\mathbf{x}[k] = \left(\bar{\mathbf{y}}[k-1],\cdots,\bar{\mathbf{y}}[k-n_v]\right)}$; and, for the ARMAX model $\mathbf{x}[k] = \left(\tilde{\mathbf{v}}[k-1],\cdots,\tilde{\mathbf{v}}[k-n_y]\right)$.

Here ${\underline{\mathbf{z}}[k] = (\underline{\mathbf{y}}[k-1], \underline{\mathbf{u}}[k])}$. Using both inputs $\underline{\mathbf{u}}[k]$ \textit{and} autoregressive terms $\underline{\mathbf{y}}[k-1]$ is what allows this state-space representation to encompass ARX, ARMAX and OE models, and also other representations such as the polynomial greybox models proposed in~\cite{noel_greybox_2018}.

\subsection{Initial conditions}
\label{sec:initial-conditions}

In order to guarantee the desirable asymptotic properties, the prediction model needs to be simulated starting with the appropriate initial conditions $\mathbf{x}_0$. Since the true initial condition $\mathbf{x}_0^*$ is unknown, there are two possible approaches when estimating the parameters.

The first approach is to fix $\mathbf{x}_0$, for some ${\mathbf{x}_0\approx \mathbf{x}_0^*}$, and minimize the cost function~(\ref{eq:cost_function}). This approach is based on the idea that,  for an asymptotically stable system, the influence of the initial conditions on the output will decrease over time for many cases of interest~\cite{boyd_fading_1985} and, hence, even if $\mathbf{x}_0\not= \mathbf{x}_0^*$ we can still obtain a good estimate of the parameters. In this case the first samples may be discarded, to make sure the transient errors are not too large. For the ARMAX model, an appropriate choice of initial values would be ${\tilde{\mathbf{v}}[k] = \mathbf{0},~k = 1, \cdots, n_v}$ and, for the OE model, ${\tilde{\mathbf{y}}[k] = \mathbf{y}[k],~k = 1, \cdots, n_y}$.

The second approach consists in including $\mathbf{x}_0$ in the optimization problem,
so it converges to $\mathbf{x}_0^*$ and improves the quality of the parameter estimates. The optimization
problem to be solved in this case is to minimize $V$ with both $\boldsymbol{\theta}$ and $\mathbf{x}_0$
as optimization variables:
\begin{equation}
  \label{eq:cost_function_ext}
  \min_{\boldsymbol{\theta}, \mathbf{x}_0} V.
\end{equation}

\section{Smoothness of prediction error methods}
\label{sec:smoothness-properties}

The theorem below relates the Lipschitz constant of $V$, and its gradient (i.e. $\beta$-smoothness), to  the simulation length~$N$. The Lipschitz constant of the cost function and the $\beta$-smoothness both play a crucial role in optimization~\cite{nesterov_introductory_1998}. Lower values imply that local (Taylor) expansions of the cost function are more predictive of the cost function, and that the optimization algorithm can still converge while taking larger steps. It also gives an upper bound on how distinct in performance two close local minima may be.

The first part of the theorem below can be seen as a formalization of the exploding gradient problem, often studied in the context of neural networks~\cite{pascanu_difficulty_2013}. The second part provides information about the explosion of second-order derivatives and curvature and it is, to the best of our knowledge, novel. As a result of the analysis, it is found that not only \textit{walls} (resulting from large first-order derivatives) might be formed in non-contractive regions of the parameter space, but also regions with exploding curvature with multiple close local minima (cf.~Fig.~\ref{fig:ms_logistic}(a)). A recurrent neural network-oriented perspective of these results is discussed in a concurrent work from our group~\cite{ribeiro_exploding_2020}.

\begin{thm}
  \label{thm:lipschitz}
  Let $\textbf{h}(\mathbf{x}, \underline{\mathbf{z}}; \boldsymbol{\theta})$ and $\textbf{g}(\mathbf{x}, \underline{\mathbf{z}}; \boldsymbol{\theta})$ in~(\ref{eq:nlss}) be  Lipschitz in ${(\mathbf{x}, \boldsymbol{\theta})}$ with constants $L_h$ and $L_g$ on a compact and convex set $\Omega = (\Omega_{\mathbf{x}},\Omega_{\underline{\mathbf{z}}}, \Omega_{\boldsymbol{\theta}})$.
  With  $\{\underline{\mathbf{z}}[k]\}_{k=1}^N \subseteq \Omega_{\underline{\mathbf{z}}}$ and $(\Omega_{\mathbf{x}}, \Omega_{\boldsymbol{\theta}})\subseteq \mathbb{R}^{N_x}\times \mathbb{R}^{N_\theta}$. If there exists at least one choice of ${(\mathbf{x}_0, \boldsymbol{\theta})}$ for which there is an invariant set contained in $\Omega$, then, for trajectories and parameters confined within  $\Omega$:
  \begin{enumerate}
      \item 
       The cost function $V$ defined in~(\ref{eq:cost_function}) is Lipschitz with constant:\footnote{Where $\mathcal{O}$ denotes the big O notation. It should be read as: $L(N) = \mathcal{O}(g(N))$ if and only if there exist positive integers $M$ and $N_0$ such that ${ |L(N)|\leq \;Mg(N)\text{ for all }N\geq N_{0}}$.}
      \begin{equation}
      \label{eq:asymptotic_L}
      L_V = 
          \begin{dcases}
          \mathcal{O}(L_h^{2N}) & \text{if } L_h > 1, \\
          \mathcal{O}(N) & \text{if } L_h = 1, \\
          \mathcal{O}(1) & \text{if } L_h < 1.
          \end{dcases}
      \end{equation}
      \item
      If the Jacobian matrices of $\textbf{h}$ and $\textbf{g}$ are also Lipschitz with respect to ${(\mathbf{x}, \boldsymbol{\theta})}$ on $\Omega$, then the gradient of the cost function $\nabla V$ is also Lipschitz with constant:
          \begin{equation}
      \label{eq:asymptotic_L'}
      L_{V}' = 
          \begin{dcases}
          \mathcal{O}(L_h^{3 N}) & \text{if } L_h > 1, \\
          \mathcal{O}\left(N^3\right) & \text{if } L_h = 1, \\
          \mathcal{O}\left(1\right) & \text{if } L_h < 1.
          \end{dcases}
      \end{equation}
  \end{enumerate}
\end{thm}

\begin{pf}
See Appendix~\ref{sec:proof_thm_ms_lipschitz}.
\end{pf}

For contractive models\footnote{We say that a dynamical system $\mathbf{x}[k+1] = \mathbf{h}(\mathbf{x}[k])$ is \textit{contractive} if, for all $\mathbf{x}$ and $\mathbf{w}$, it satisfies $\|\mathbf{h}(\mathbf{x}) - \mathbf{h}(\mathbf{w})\| < L \|\mathbf{x} -\mathbf{w}\|$, for $L < 1$.}, under certain regularity conditions, we have $L_h<1$ and, according to the above theorem, both the Lipschitz constant and the $\beta$-smoothness of the cost function can be bounded by a constant that, asymptotically, does not depend on the simulation length. All contractive systems have a unique fixed point inside the contractive region, and all trajectories converge to such a fixed point~\cite[Theorem 9.23]{rudin_principles_1964}. Systems with richer nonlinear dynamic behaviors, such as limit cycles and chaotic attractors, and also unstable systems, are \textit{non-contractive} and will always have $L_h\ge1$. The Lipschitz constants and $\beta$-smoothness  for these systems may, according to Theorem~\ref{thm:lipschitz}, blow up exponentially (or polynomially for some limit cases) with the maximum simulation length.

Less formally, for models that have infinitely long dependencies (i.e.~are non-contractive) the distance between trajectories of models that are close in the parameter space might become progressively larger along the simulation length because errors will accumulate. This might yield very intricate objective functions in some parts of the parameter space  and make the optimization problem very dependent on the initial estimate.

\section{Multiple shooting}
\label{sec:multiple-shooting}

\begin{figure*}[t]
  \centering
  \subfloat[]{\includegraphics{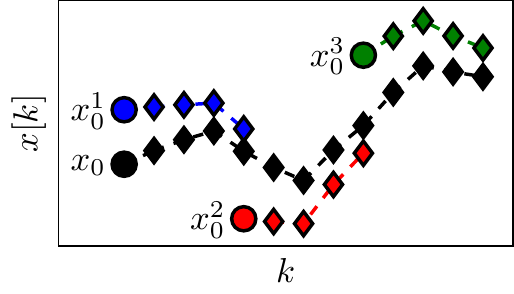}}
  \subfloat[]{\includegraphics{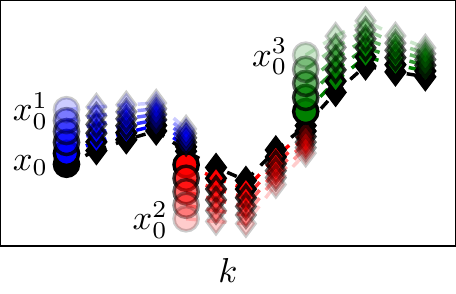}}
  \subfloat[]{\includegraphics{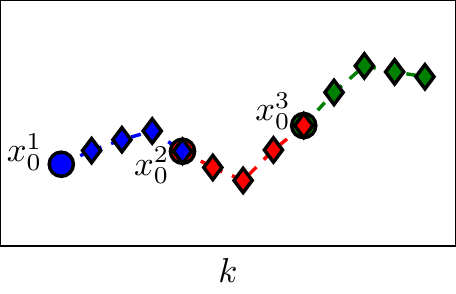}}
  \caption{\textbf{Multiple shooting illustration}.  In \textbf{black}, we present the simulation of the dynamic system through the entire window length using the single initial condition $x_0$ (represented by \protect\includegraphics{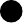}). The simulated values are represented by \protect\includegraphics{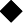}. Dividing the window length into three sub-intervals and simulating the system in each of these, for initial conditions $x_0^1$, \protect\includegraphics{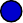}, $x_0^2$, \protect\includegraphics{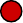}, and $x_0^3$, \protect\includegraphics{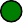}, results in the three different simulations represented by \protect\includegraphics{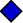}, \protect\includegraphics{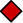}~and \protect\includegraphics{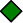}, respectively. In (a), the end of one simulation does not coincide with the beginning of the next one ($x^{i-1}[{\rm m}_i] \not= x_0^i$). In (b), we show what happens as $\|x^{i-1}[{\rm m}_i] - x_0^i\| \rightarrow 0$. And, in (c), we show that, when $x^{i-1}[{\rm m}_i] = x_0^i$, the concatenation of these short simulations is equivalent to a single one carried out over the entire window length.}
  \label{fig:ms_ilustration}
\end{figure*}

The theoretical results from the previous section suggest that long simulation lengths might yield regions of the parameter space where the cost function is intricate, hence hard for the optimization algorithm to navigate. 

In this section, we propose the application, in the context of prediction error methods, of a technique called \textit{multiple shooting} for which the maximum simulation length is a design parameter. This enables solving problems that would be impossible or very hard to solve in the setting of Section~\ref{sec:pred-error-meth}, which will be named \textit{single shooting} from now on.

\subsection{Method formulation}

For the \textit{multiple shooting} formulation, rather than simulating the prediction model~(\ref{eq:nlss}) through the entire dataset from a single initial condition vector $\mathbf{x}_0$, the data is split into $M$ intervals 
${\{[{\rm m}_i +1, {\rm m}_{i+1}] \mid i = 1, \cdots, M\}}$, 
$0 = {\rm m}_1 < {\rm m}_2 < \cdots < {\rm m}_M < {\rm m}_{M+1} = N$,
each one with its own set of initial conditions $\mathbf{x}_{0}^i\in \mathbb{R}^{N_x}$.
The $i$-th vector of initial conditions $\mathbf{x}_{0}^i$ is used to compute 
$\hat{\mathbf{y}}^i[k]$ over ${{\rm m}_i+1 \le k  \le {\rm m}_{i+1}}$:
\begin{subequations}
  \label{eq:ms}
  \begin{align}
  \mathbf{x}^i[k] &= \mathbf{h}(\mathbf{x}^i[k-1], \underline{\mathbf{z}}[k]; \boldsymbol{\theta}), \text{ for }\mathbf{x}^i[{\rm m}_i] = \mathbf{x}_{0}^i,\\
  \hat{\mathbf{y}}^i[k] &= \mathbf{g}(\mathbf{x}^i[k], \underline{\mathbf{z}}[k]; \boldsymbol{\theta}).
  \end{align}
\end{subequations}
Since the length of the simulation is limited to the shorter interval $[{\rm m}_i +1,  {\rm m}_{i+1}]$, the trajectory is less likely to strongly diverge and this typically helps the optimization procedure by making the objective function \textit{smoother}.

Let ${\Delta {\rm m}_i = {\rm m}_{i+1} - {\rm m}_i}$, we define:
\begin{equation}
  \label{eq:Vi}
  V_i = \frac{1}{\Delta {\rm m}_i}\sum_{k={\rm m}_i+1}^{{\rm m}_{i+1}} \|\mathbf{y}[k] - \hat{\mathbf{y}}^i[k]\|^2,  
\end{equation}
to be the cost function associated with the $i$-th interval, where the prediction $\hat{\mathbf{y}}^i[k]$ depends upon $\boldsymbol{\theta}$ and $\mathbf{x}_{0}^i$, according to~(\ref{eq:ms}). The \textit{multiple shooting} formulation makes use of the following objective function:
\begin{equation}
    \label{eq:cost_function_ms}
    V^{M} = \sum_{i=1}^M \tfrac{\Delta {\rm m}_i}{N} ~ V_{i} .
\end{equation}
This objective function includes states $\mathbf{x}_0^1,  \cdots,  \mathbf{x}_0^M$  as free variables in the optimization. Hence, rather than reinforcing the cohesion of the states $\mathbf{x}[k]$
by defining them through a recurrence relation that casts a dependency
of $\mathbf{x}[k]$ all the way back to the initial condition $\mathbf{x}_0$,
as in the \textit{single shooting} formulation, the cohesion between subsequent states is
achieved through optimization constraints, resulting in the following problem:
\begin{align}
\label{eq:opt_prob_ms}
  \min_{\boldsymbol{\theta},  \mathbf{x}_0^1,  \cdots,  \mathbf{x}_0^M} &   V^{M}, \\\nonumber
  \text{subject to: } & \mathbf{x}^{i-1}[{\rm m}_i] = \mathbf{x}_0^i,\\\nonumber &\text{for } i=2, 3, \cdots, M .
\end{align}
The next theorem provides the equivalence between~(\ref{eq:cost_function_ext}) and~(\ref{eq:cost_function_ms}). The theorem and its corollary are a formalization of the intuition provided in Fig.~\ref{fig:ms_ilustration} and they give further insight into how the constraints in the \textit{multiple shooting} formulation are used to imitate a single simulation throughout the entire dataset.

\begin{thm}
  \label{thm:ms_equivalence_cost}
  If $\mathbf{x}^{i-1}[{\rm m}_i] = \mathbf{x}_0^i, \text{for } i=2, 3, \cdots, M$ and $\mathbf{x}_0^1 = \mathbf{x}_0$, then $V =  V^M$.
\end{thm}
\begin{pf}
Let us call $\mathbf{x}[k]$, $\hat{\mathbf{y}}[k]$ and $\mathbf{x}^{i}[k]$, $\hat{\mathbf{y}}^i[k]$  the states and predictions in, respectively, the single shooting simulation and in the $i$-th multiple shooting interval. For a fixed $i$, if $\mathbf{x}[{\rm m}_i] = \mathbf{x}_0^i$ then $\mathbf{x}[k] = \mathbf{x}^{i}[k]$ for all ${k\in [{\rm m}_i+1, {\rm m}_{i+1}]}$.  Hence, inside this same interval, $\hat{\mathbf{y}}[k] = \hat{\mathbf{y}}^i[k]$. Applying this for every $i$ it follows from the respective definitions that: $V = \sum_{i=1}^M \tfrac{\Delta m_i}{N} ~ V_{i} =  V^M$.
\end{pf}

\begin{cor}
  \label{thm:ms_equivalence_solution}
  The pair $(\boldsymbol{\theta}^*, \mathbf{x}_0^*)$ is a global solution
  of~(\ref{eq:cost_function_ext}) if and only if there exist $(\mathbf{x}_0^2, \cdots, \mathbf{x}_0^M)$
  such that $(\boldsymbol{\theta}^*, \mathbf{x}_0^*, \mathbf{x}_0^2, \cdots, \mathbf{x}_0^M)$
  is a global solution of the optimization problem~(\ref{eq:opt_prob_ms}).
\end{cor}

Multiple shooting can be understood as a generalization of the single shooting case. That is because, if $M=1$ and $\Delta {\rm m}_{1}=N$, both methods result in the same optimization problem. Multiple shooting, however, might provide some advantages. The method is more amenable to parallelization, since each cost function $V_i$ and its associated  derivatives can be computed independently and, possibly, in parallel. Also, long simulations usually yield larger numerical errors due to finite precision errors that accumulate along the simulation, hence multiple shooting shorter simulation intervals also attenuate this problem. Finally, multiple shooting cost function has better smoothness properties that will be investigated next.

\subsection{Properties of the cost function}
\label{sec:objective_function_properties}

For the multiple shooting method, the Lipschitz constant of $V^M$ and of its gradient, $L_{V^M}$ and $L_{V^M}'$ depend asymptotically on $\Delta \text{m}_{\max} = \max_{1\le i \le M} \Delta \text{m}_i$ rather than  on $N$ (See Appendix~\ref{sec:lipsch-analys-mult}). For instance, if $L_h > 1$: 
\begin{equation}
L_{V^M} = \mathcal{O}(L_h^{2\Delta \rm{m}_{\max}}); ~~
L_{V^M}' =  \mathcal{O}(L_h^{3 \Delta \rm{m}_{\max}}).
\end{equation}
Since $\Delta \rm{m}_{\max}$ is a design parameter, it is possible to have some control over the Lipschitzness and $\beta$-smoothness of the objective function in the non-contractive  region of the parameter space ($L_h \ge 1$).

\section{Implementation and numerical examples}
\label{sec:impl-numer-exampl}

In this section, numerical examples are presented. The cost function smoothness is investigated through the lens of the theoretical results in Section~\ref{sec:smoothness-properties} and it is shown how \textit{multiple-shooting} might help mitigate some problems.

The equality-constrained problem~(\ref{eq:cost_function_ms}), which arises from the multiple shooting formulation, is solved using an implementation of the sequential quadratic programming solver originally described in~\cite{lalee_implementation_1998} available in the SciPy library~\cite{virtanen_scipy_2020}\footnote{\texttt{ scipy.optimize.minimize(method=`trust-constr')}}. The procedure used for computing the derivatives is explained in Appendix~\ref{sec:comp-deriv}. Additional numerical examples are provided in Appendix~\ref{sec:additional-experiments}.

\subsection{OE model for a chaotic system}
\label{sec:output-error-chaotic-system}

This example illustrates how multiple shooting makes prediction error methods more robust w.r.t. the choice of initial conditions for the optimization. A dataset with $N=200$ samples is generated using the logistic map~\cite{may_simple_1976}:
\begin{equation}
  \label{eq:logistic_map}
  y[k] = \theta y[k-1](1-y[k-1]),
\end{equation}
with  $\theta=3.78$. From the generated dataset we try to estimate an \textit{output error} model with the same structure.

\begin{figure*}
  \centering
  \subfloat[$\Delta \rm{m}_{\max}= N$ (Single Shooting)]{\includegraphics{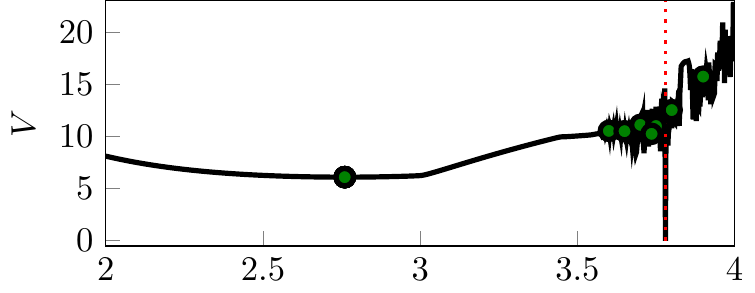}}
  \subfloat[$\Delta \rm{m}_{\max} = 10$]{\includegraphics{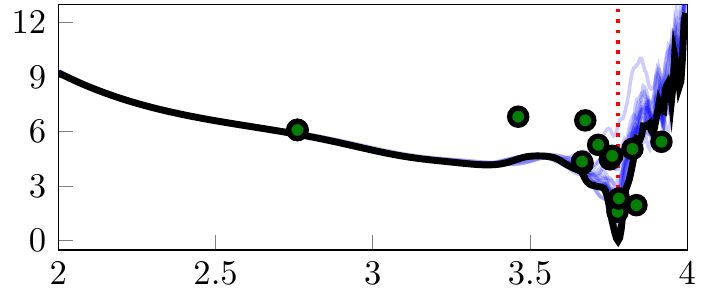}}\\
  \subfloat[$\Delta \rm{m}_{\max} = 5$]{\includegraphics{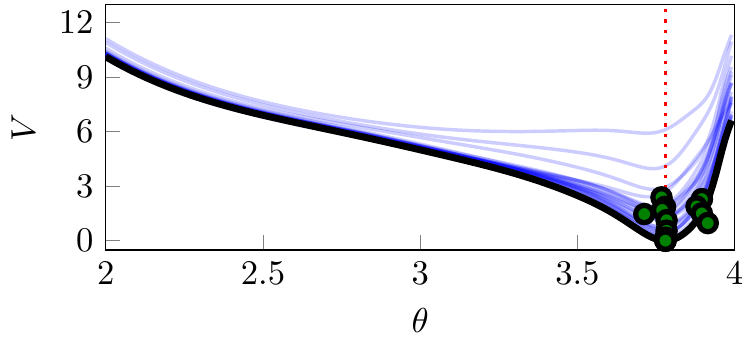}}
  \subfloat[$\Delta \rm{m}_{\max} = 2$]{\includegraphics{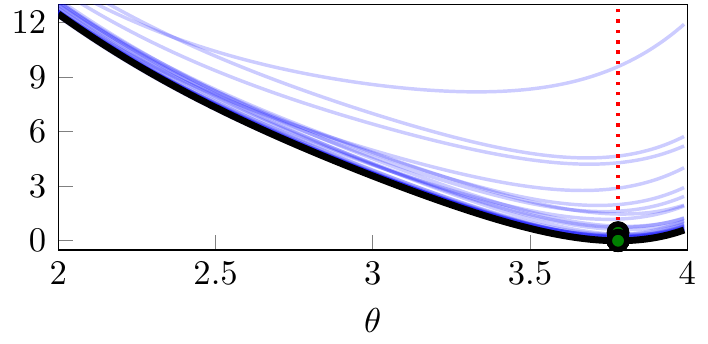}}
  \caption{\textbf{Logistic map parameter estimation}. \textit{Cost function} of the optimization problem for $\mathbf{x}_0^i$ fixed in its true values (in black) and for disturbed versions of the these initial conditions (in blue). We present the result for four values of $\Delta \rm{m}_{\max}$, and omit disturbed initial condition objective functions for the single shooting case ($\Delta \rm{m}_{\max}= N$) to make it easier to visualize. The green circles, \protect\includegraphics{img/greencircle.pdf}, indicate the pair $(\theta, V)$ corresponding to a solution found by the solver. There are 15 circles in each figure (some of them overlapping), the  circles correspond to solutions for different initial guesses. As initial guesses we picked values of $\theta$ uniformly spaced between 3.2 and 3.9, with $\mathbf{x}_0^i$ picked from  randomly disturbed versions of the true initial conditions (which are known because we generated the data ourselves). The true value $\theta=3.78$ is indicated by the dotted red vertical line.}
  \label{fig:ms_logistic}
\end{figure*}

Fig.~\ref{fig:ms_logistic}(a) illustrates the objective function for the \textit{single shooting} case. The data generating system, Eq.~(\ref{eq:logistic_map}), presents chaotic behavior for $\theta\in [3.57, 4]$, which explains the very intricate objective function in this region. For chaotic systems, small variations in the parameters may cause large variations in the system trajectory and, consequently, abrupt changes in the \textit{free-run simulation error}. This explains why the cost function used in estimating an OE model has many local minima for this problem.

The solutions found by the solver, for different initial guesses, are also displayed in Fig.~\ref{fig:ms_logistic}(a). Notice that the solver fails to find the true solution because it always gets trapped in local stationary points near the initial guess. 
Even in this noiseless scenario, the estimation problem is very challenging due to the chaotic nature of the system. Hence, for a simulation that is sufficiently long, the trajectories will differ significantly even for small parameter variations.

Multiple shooting makes the problem easier by limiting the maximum simulation length. Fig.~\ref{fig:ms_logistic} (b), (c) and (d) display the objective function and the solutions found by the solver starting from different initial guesses. The estimation procedure becomes easier as $\Delta \rm{m}_{\max}$ is made smaller. For Fig.~\ref{fig:ms_logistic}(c) the solver already converges to the true parameter for some initial guesses,  but not for all of them. For Fig.~\ref{fig:ms_logistic}(d), the solver converges to the true solution for all initial guesses that have been tested.

For the multiple shooting case, besides $\theta$, the initial conditions are also optimization parameters. To help with the visualization of this multidimensional problem, Fig.~\ref{fig:ms_logistic}(b), (c) and (d)  display the main curve corresponding to the objective function for the true initial conditions and faded lines corresponding to the objective function for perturbed initial conditions.  Another consequence of the problem having more parameters than displayed in the figure is that the cost function found by the solver does not need to lie on any of the objective function curves displayed in the figure, since it may have a different set of initial conditions $\mathbf{x}_0^i$.

It is important to highlight that the mechanism used here is not to take the system outside of the chaotic regime, but rather avoid simulating the system for too long. By doing that, we avoid the major problem that arises in the identification of chaotic systems, i.e. the high sensitivity to parameters and initial conditions. This results in a better behaved objective function (cf. Fig.~\ref{fig:ms_logistic}). The constraints allow the equivalence with the original prediction error problem (according to Theorem~\ref{thm:ms_equivalence_cost} and Corollary~\ref{thm:ms_equivalence_solution}).

Table~\ref{tab:performance_logistic} gives the number of function evaluations and the running time for the four situations displayed in Fig.~\ref{fig:ms_logistic}. The convergence happens within just a few iterations for $\Delta {\rm m}_{\max} = N$ (single shooting) because any initial point is probably very close to some optimal \textit{local} solution. As we reduce $\Delta \rm{m}_{\max}$ the objective function becomes less intricate and this is reflected in the convergence of the solver. For $\Delta \rm{m}_{\max} = 10$ the solver takes much longer to converge. We believe  this happens because the local solution is not so close in the parameter space to the initial guess anymore. As $\Delta \rm{m}_{\max}$ is further decreased, however, the convergence becomes faster because it is dealing with, what is believed to be, a smoother problem that can be handled more accurately by low order approximations.
\begin{table}[htpb]
  \setlength\extrarowheight{2pt}
  \centering
  \caption{\textbf{Computational cost for the logistic map estimation.}
    The number of function evaluations and total running time until convergence. 
    Minimum, maximum and median are given over
     15 runs for the situations presented in
     Fig.~\ref{fig:ms_logistic}. *The number of iterations
     is limited to 1000 and the solver is interrupted when this number is reached.}
  \begin {tabular}{c|ccc|ccc}
    \multicolumn{1}{c|}{} &\multicolumn{3}{c|}{function evaluations} & \multicolumn{3}{c}{run time (s)}\\
    \hline
    $\Delta \rm{m}_{\max}$ & min & median & max &  min & median & max  \\\hline
    $N$ & 1 & 15 & 23 & 0.01 & 0.2 & 1.1  \\ \hline 
    $10$ & 43 & 1000* & 1000* & 0.7 & 24.7 & 27.3 \\ \hline
    $5$ & 29 & 115 & 645 & 1.0 & 2.9 & 26.9 \\ \hline
    $2$ & 21 & 50 & 65 & 1.7 & 2.9 & 3.8 \\ \hline
  \end{tabular}
  \label{tab:performance_logistic}
\end{table}

\begin{figure*}[t]
  \centering
\subfloat[]{\includegraphics{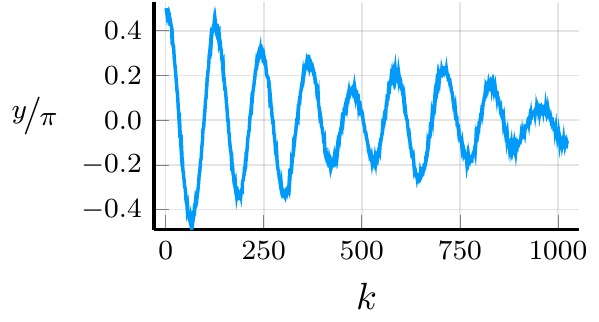}}
\subfloat[]{\includegraphics{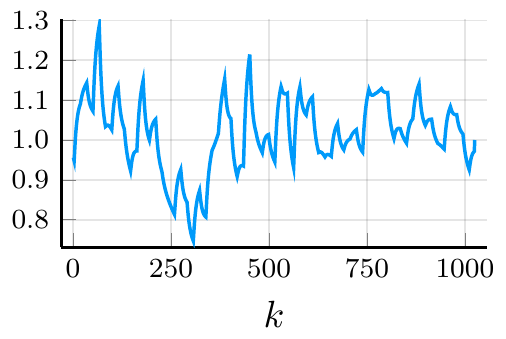}}
\subfloat[]{\includegraphics{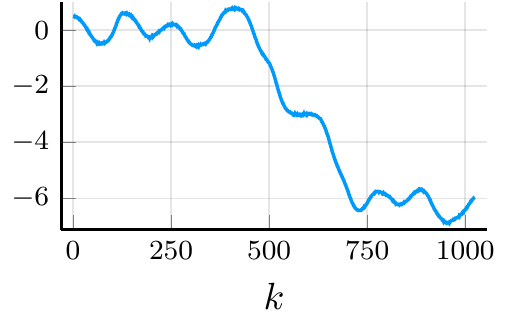}}
\caption{\textbf{Angle over time for pendulum parameter estimation}. Display the \textit{output signal} $y[k]$ for the three different datasets used for estimating the parameters. The dataset size is $N=1024$ samples. (a) The input applied in this case is a zero-mean Gaussian random input with standard deviation $\sigma_u = 10$, each random value being held for $20$ samples. The input in this case is unable to drive the system away from the influence of the stable fixed point $(0, 0)$. (b) The input $u[k]$ in this case is obtained by the control law: $u[k] = 40\delta e[k-1] - 78.8\delta e[k-2] + 38.808\delta e[k-3] + 1.02 u[k-1] -0.02 u[k-2],$ where  the error is the difference between the reference and the output: $e[k] = r[k] - y[k]$. The reference is $r[k] = \pi + \Delta r[k]$ where $\Delta r[k]$  is a zero-mean Gaussian random input with standard deviation $\sigma_r = 0.2$, each random value being held for $20$ samples; and, (c) same as (a) but with the larger standard deviation $\sigma_u = 50$, which is able to drive the pendulum to complete full rotations. For (a) and (c) zero-mean Gaussian white noise with standard deviation $\sigma_r = 0.03$ was added to the output.}
\label{fig:output_signal_pendulum}
\end{figure*}

\subsection{Pendulum and inverted pendulum}
\label{sec:pendulum-and-inverted-pendulum}

Consider the following discrete-time nonlinear system:
\begin{eqnarray}
&&\left\{
\begin{array}{l}
x_1[k+1] = x_1[k] + \delta x_2[k]\\
x_2[k+1] = -\delta \frac{g}{l} \sin x_1[k] +(1-\delta\frac{k_a}{m}) x_2[k] + \delta \frac{1}{m} u[k]
\end{array}
\right.\nonumber\\
&&
y[k] = x_1[k] \label{eq:pendulum_eq}
\end{eqnarray}

\noindent
which corresponds to a pendulum model, discretized using the Euler approximation ${\dot{x}(t) \approx \frac{x((k+1)\delta) - x(k\delta)}{\delta}}$, where $g$ is the gravity acceleration, $m$ is the mass connected to the extremity of the pendulum, $l$ is the length of the (massless) rod connecting the mass to the pivot point, and $k_a$ is the linear friction constant. It has two states: the angle of the mass ($x_1$) and the angular velocity ($x_2$). The input $u[k]$ is the force applied to the mass.

This system has multiple equilibrium points, namely, $(x_1, x_2) = (\pm\pi i, 0)$ for $i = 0, 1, 2, 3, \cdots$. The equilibrium points at $(x_1, x_2) = (\pm 2 \pi i , 0)$ are stable and the equilibrium points at $(x_1, x_2) = (\pi \pm 2 \pi i , 0)$ are unstable. For this system, with $g = 9.8$, $l = 0.3$, $m = 3$, $k_a = 2$ and $\delta = 0.01$, we define three different datasets: 
(a)~A dataset for which small inputs are applied to the system, that stays under the influence of the stable point $(x_1, x_2) = (0, 0)$ and $y[k]$ stays, approximately, inside the range $\left[-\frac{\pi}{2}, +\frac{\pi}{2}\right]$; (b)~A dataset for which the system is maintained close to the unstable point $(x_1, x_2) =(\pi, 0)$  by a linear controller; and, (c)~A dataset for which the input is large enough to drive the pendulum to full rotations around its center. The output  corresponding to those three situations are displayed in Fig.~\ref{fig:output_signal_pendulum}.

\begin{figure*}
  \centering
\subfloat[]{\includegraphics{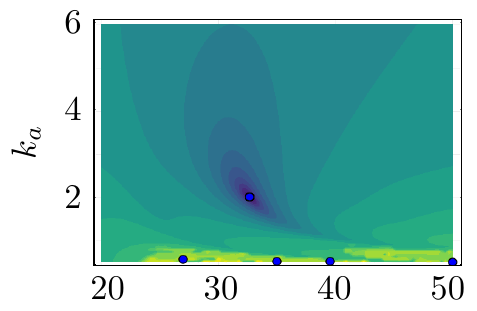}}\hspace{0.7cm}
\subfloat[]{\includegraphics{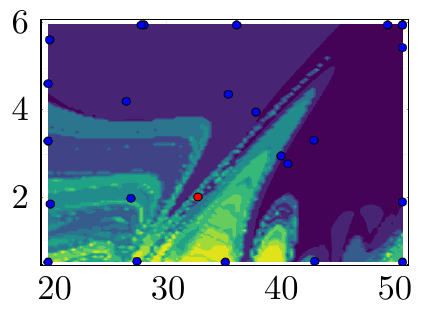}}
\subfloat[]{\hspace{0.7cm}\includegraphics{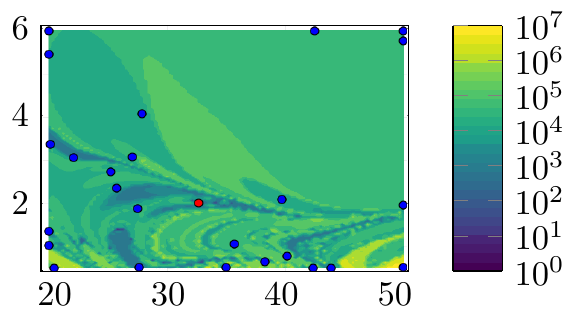}}
\\
\subfloat[]{\includegraphics{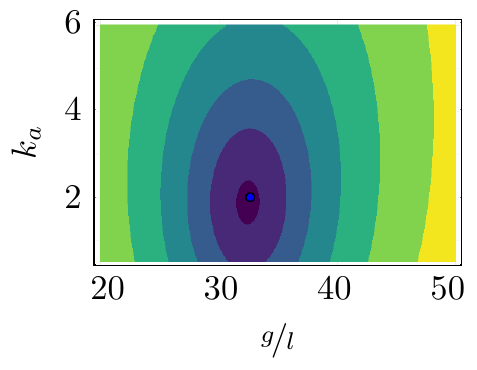}}\hspace{0.7cm}
\subfloat[]{\includegraphics{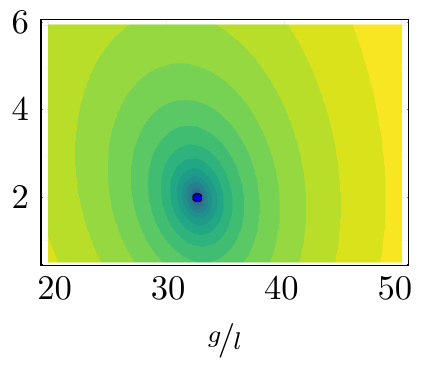}}
\subfloat[]{\hspace{0.7cm}\includegraphics{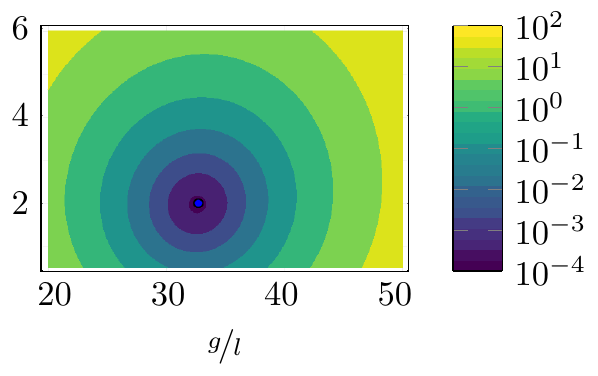}}
\caption{\textbf{Contour plot of the cost function.} Figures (a), (b) and (c) correspond to the cost function $V$ from single shooting simulation for the datasets (a), (b) and (c) generated as described in Fig.~\ref{fig:output_signal_pendulum} caption; and, in (d), (e) and (f) the cost function for the same problems is displayed for the multiple shooting formulation with $\Delta \rm{m}_{\max}=16$. The true parameter is indicated by a red circle, \protect\includegraphics{img/redcircle.pdf}, and the solutions found by the solver are indicated by blue circles, \protect\includegraphics{img/bluecircle.pdf}. There are 25 blue circles in each figure (some of them overlapping with each other), each circle corresponds to the solution for a different initial guess. The red circles might not be visible in some cases because they are hidden under the blue circles. Some solutions are outside of the displayed region and the corresponding blue dots are displayed at the edge of the plot. Initial optimization guesses for $\theta$ were picked from a grid of points uniformly spaced in the rectangle $[20, 50] \times [0.5, 6]$. It is important to highlight that the plots show a two-dimensional projection of a cost function that is defined on an extended parameter space that includes the initial conditions $\mathbf{x}_{0}^i$, $i = 1, \cdots, M$ as parameters, which were fixed to the true values when generating the contour plots.}
\label{fig:cost_function_pendulum}
\end{figure*}

Fixing $m = 3$ and $\delta = 0.01$ parameters $\frac{g}{l}$ and $k_a$ of an output error model with the structure presented in~(\ref{eq:pendulum_eq}) were estimated from the data. A visualization of the cost function is presented in Fig.~\ref{fig:cost_function_pendulum} together with numerical solutions found by means of the single shooting and multiple shooting formulation starting from different initial conditions.

For dataset (a), the single shooting formulation is able to recover the true parameters from data for most of the initial conditions. Some exceptions occur when initialized far away from the correct initial conditions. For datasets (b) and (c), for which the system needs, respectively, to operate close to the unstable dynamics or to account for the existence of multiple fixed points, the cost function is highly intricate and full of local minima. In this case, the optimization algorithm, even when initialized close to the local solution, fails to converge to reasonable solutions. This result is consistent with Theorem~\ref{thm:lipschitz} and how the smoothness of the objective function degenerates (exponentially) on sets of the parameter space for which the prediction model is non-contractive, such as the trajectories close to the unstable fixed point of the system~(\ref{eq:pendulum_eq}). The use of multiple shooting yields an objective function that looks similar to a paraboloid in the region of interest for the three cases and the optimization procedure converges to the true parameter regardless of the initialization.

For nonlinear ARX, ARMAX or OE models the states are directly measured (possibly with some noise contamination) and the initialization of the optimization parameters $\mathbf{x}_0^i$  follows naturally from the considerations in Section~\ref{sec:initial-conditions}. Here, however, the state variable $x_2$ is not measured. This variable can be interpreted as the derivative of $x_1$, so a finite difference approximation was used in the initialization of the intermediary initial conditions $\mathbf{x}_0^i$. Although the high-pass behavior of the derivative amplifies the noise, this choice was still better than a completely arbitrary one.

\section{Comparison with multi-step-ahead prediction error minimization}
\label{sec:comparison}

\subsection{Multi-step-ahead prediction error minimization}

Multiple-shooting is presented here as a possible way of limiting the simulation interval~$\Delta \rm{m}_{\max}$. A method that appears in the system identification literature that also has a similar effect is the multi-step-ahead prediction error minimization (MSA-PEM)~\cite{farina_simulation_2011},~\cite{terzi_learning_2018}~\cite{farina_identification_2012}. The approach  fits well into the moving horizon framework~\cite{terzi_learning_2018} and is popular for system identification in model predictive control application.  The method has been studied primarily in a linear model setting, nevertheless it can be extended to a nonlinear setting~\cite{farina_identification_2012}. 

In the MSA-PEM estimation, the simulation is truncated to a fixed number $K$ of steps backwards. That is, for each $k = 1, \cdots, N$, we  define an auxiliary variable $\tilde{\mathbf{x}}_k[i]$ and an initial condition $\tilde{\mathbf{x}}_{0, k}$. Starting from $\tilde{\mathbf{x}}_k[k-K] = \tilde{\mathbf{x}}_{0, k}$ the system is propagated using the state equation: $\tilde{\mathbf{x}}_k[i] = \mathbf{h}(\tilde{\mathbf{x}}_k[i-1], \underline{\mathbf{z}}[i]; \boldsymbol{\theta})$ to simulate the evolution of this auxiliary state variable for $i=k-K, \cdots, k$. The prediction is then computed using $\hat{\mathbf{y}}[k] = \mathbf{g}(\tilde{\mathbf{x}}_k[k], \underline{\mathbf{z}}[k]; \boldsymbol{\theta})$. The parameters are obtained by minimizing a cost function similar to~(\ref{eq:cost_function}).

Multiple shooting is equivalent, in the sense of Theorem~\ref{thm:ms_equivalence_cost} and Corollary~\ref{thm:ms_equivalence_solution}, to solving the original (single shooting) problem regardless of the choice of simulation interval $\Delta m_{\text{max}}$. MSA-PEM, on the other hand, is equivalent to the original formulation only if $K = N$. The next example illustrates how the statistical properties of the method change as $K$ varies from $1$ to $N$.

\begin{figure*}[t]
  \centering\hspace{0.01cm}
  \subfloat[][]{\includegraphics{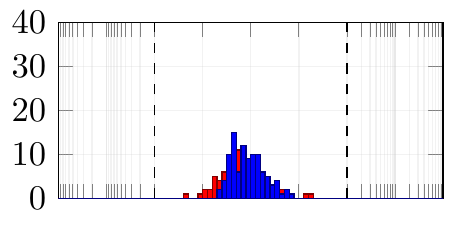}}\hspace{0.05cm}
  \subfloat[][]{\includegraphics{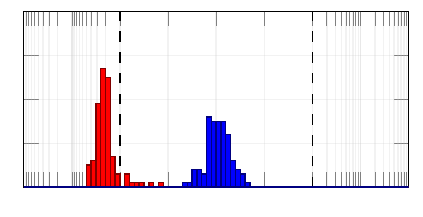}}\hspace{0.05cm}
  \subfloat[][]{\includegraphics{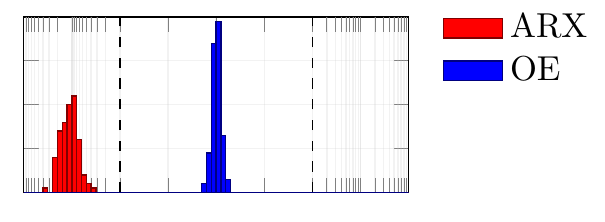}}\\\hspace{0.25cm}
  \subfloat[][]{\includegraphics{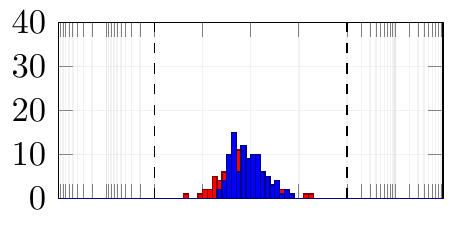}}\hspace{0.05cm}
  \subfloat[][]{\includegraphics{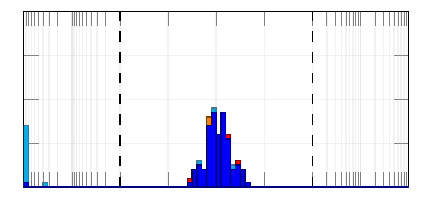}}\hspace{0.05cm}
  \subfloat[][]{\includegraphics{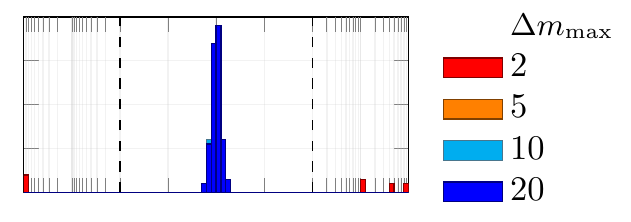}}\\
  \subfloat[][]{\includegraphics{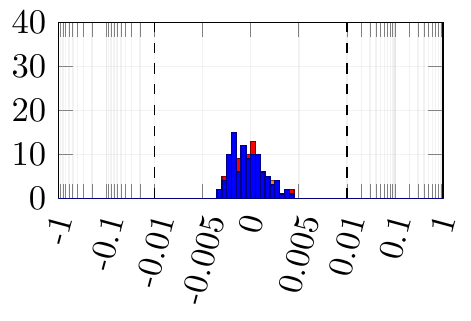}}
  \subfloat[][]{\includegraphics{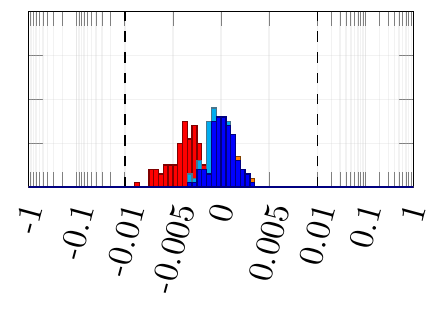}}
  \subfloat[][]{\includegraphics{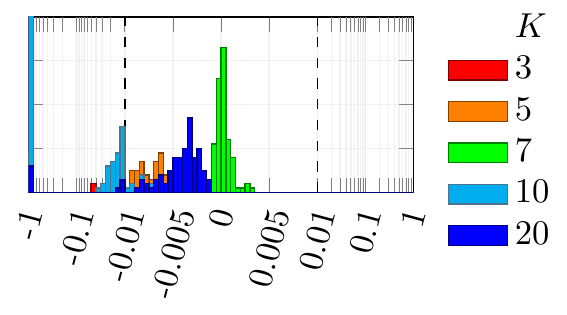}}
\caption{\textbf{Monte Carlo simulation of the estimates.}  \textit{Histogram} showing $(\hat{\theta}_1-\theta_1)$, for 100 different realizations of the estimation experiment. In (a) to (c),
results for ARX and the standard (single shooting) output error estimation. The results are displayed: in (a) for the fast-response system with $\tau = 0.25$; in (b) for the intermediate setting $\tau = 0.75$; and, in (c) the slow-response system with $\tau = 0.9$.  In (d) to (f), results for multiple-shooting estimation in the same three settings. The results are displayed for $\Delta \rm{m}_{\max} = \{2, 5, 10, 20\}$, which often overlap. In (g) to (i),  results for the multi-step ahead prediction in the three settings. We show the results for $K = \{3, 5, 7, 10, 20\}$. In order to facilitate the visualization two different scales were used for the $x$-axis: between -0.01 and 0.01 the scale is linear; and logarithmic scale is used for values below -0.01 or above 0.01.
}
\label{fig:histogram_ex4}
\end{figure*}

\subsection{Example: estimating the output error model for second-order under-damped system}
\label{sec:exampl-estim-outp}

A dataset with $N = 300$ samples is generated using the equation:
\begin{subequations}
\begin{align}
\bar{y}[k] &= \theta_1 \bar{y}[k] + \theta_2 \bar{y}[k-2] +  \theta_3 u[k-1] \\
y[k] &= \bar{y}[k] + v[k].
\end{align}
\end{subequations}
Here $\bar{y}$ represents the noiseless output, and $v$ represents the white output noise that is introduced during the data generation. At each $k$, $v[k]$ is a Gaussian random variable with standard deviation $\sigma_v = 0.05$.

\begin{figure}[t]
  \centering
  \subfloat[][MS]{\includegraphics{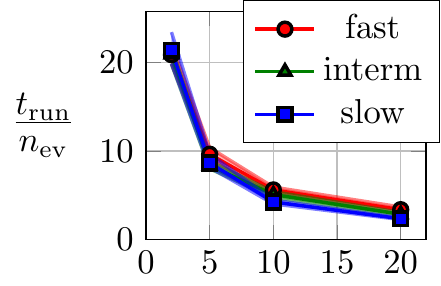}}
  \subfloat[][MSA-PEM]{\includegraphics{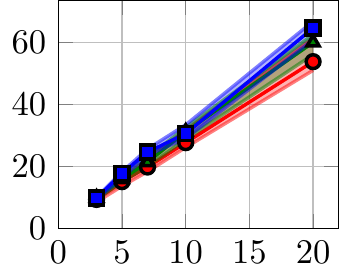}}\\
  \subfloat[][MS]{\hspace{0.2cm}\includegraphics{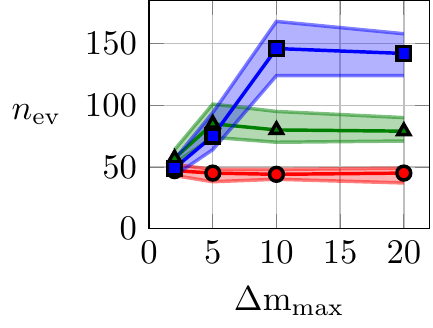}}
  \subfloat[][MSA-PEM]{\includegraphics{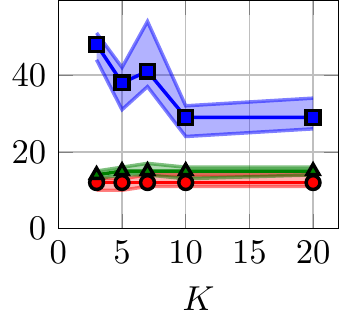}}
  \caption{\textbf{Execution time comparison}. Two components of the time for the optimization algorithm to converge are displayed: plots (a) and (b), display the \textit{total running time} $t_{\text{run}}$ divided by the number of function evaluations $n_{\text{ev}}$; and, plots (c) and (d), display the \textit{number of function evaluation}s $n_{\text{ev}}$. In (a) and (c), the results are for multiple shooting estimation and the $x$-axis is the maximum simulation length $\Delta {\rm m}_{\max}$. In (b) and (d), the results are for MSA-PEM and  the $x$-axis is the number of steps ahead $K$. The results are displayed for: the fast-response system with $\tau = 0.25$ in \textcolor{red}{red}; the intermediate setting $\tau = 0.75$ in \textcolor{green_new}{green}; and, the slow-response system with $\tau = 0.9$ in \textcolor{blue}{blue}. The results are for the 100 different realizations of experiment: the solid line is the median and the shaded region is the interquartile range. That is, the shaded region illustrate the statistical dispersion of the observed values and show the region between the first and third quartiles (i.e. 25th and 75th percentiles).}
\label{fig:computational_time_ex4}
\end{figure}

Let $\boldsymbol{\theta} = (\theta_1, \theta_2, \theta_3)$ denote the parameter vector used for the data generation. Three different settings are considered: (a)  $\boldsymbol{\theta}= (0.5, -0.2, 2)$; (b) $\boldsymbol{\theta}  = (1.5, -0.7, 0.5)$; and, (c) $\boldsymbol{\theta}  = (1.8, -0.95, 0.1)$.  The three settings correspond to underdamped linear systems with different response-times: in (a) the system responds \textit{fast} to input changes and has a time constant $\tau = 0.25$; in (c), it responds \textit{slowly} and $\tau = 0.9$; and, (b) is an \textit{intermediate} setting with $\tau = 0.75$.

From the synthetically generated data, the parameters of a model with the same structure are estimated. The experiment is repeated 100 times, each time corresponding to a different realization (i.e., different random seed) of the data generation. The result of the Monte Carlo procedure is displayed in Fig.~\ref{fig:histogram_ex4} for the parameter estimation using ARX, standard (single shooting) output error, multiple shooting, and MSA-PEM. We show the difference between estimated and true values for the first parameter, $\hat{\theta_1} - \theta_1$. Similar results have been obtained for $\hat{\theta}_2$ and $\hat{\theta}_3$.

Due to the presence of white output error, using \textit{output error estimation} yields consistent and well-behaved results. \textit{ARX estimation}, on the other hand, is biased, and its histogram is not centered around zero. This is more evident either in~\ref{fig:histogram_ex4}(b) or in~\ref{fig:histogram_ex4}(c). In Fig.~\ref{fig:histogram_ex4}(a), which correspond to the fast response system, there is not much difference between the two estimation procedures, mostly because any disturbance fade away very quickly, yielding similar statistical properties for the two estimators.

\textit{Multiple shooting estimation} results are similar to the (single shooting) output error estimation. For some choices of $\Delta \rm{m}_{\max}$, the optimization problem might actually be made harder due to the increase in the problem dimension. In this case, the algorithm seems to fail a few times, which can be observed in the histogram as a few outliers in the estimated parameter distribution. Except for these outliers, all the different choices of $\Delta \rm{m}_{\max}$ yield very similar distributions for the estimated parameters.

For \textit{MSA-PEM}, the estimation results vary qualitatively among the three settings.  For the intermediate setting, midway qualitative behavior between ARX and output error estimation is obtained, approaching that of output error estimation as $K$ increases. For the fast-response system, a similar interpretation is possible, however, there is not much difference between using ARX and output error estimation to begin with (cf. Fig~\ref{fig:histogram_ex4} (a)), hence all $K$-step-ahead choices yield similar results. Finally, for the slow-response setting, minimizing the multi-step-ahead prediction error does not offer a direct compromise between ARX and output error estimation and, for $K =10$ or $K =20$, it is possible to observe a bimodal distribution with one peak distant from zero. And $K = 7$  yields the best results.

In Fig.~\ref{fig:computational_time_ex4}, the computational cost of multiple shooting and MSA-PEM are compared. The average \textit{running time divided by the number of function evaluations} is displayed in Fig.~\ref{fig:computational_time_ex4}(a) and (b). This accounts for the computational cost of computing the cost function, derivatives, and performing the matrix factorizations needed by the optimization algorithm at each iteration. Figure~\ref{fig:computational_time_ex4}(c) and (d) display the number of cost function evaluations needed for the optimization algorithm to converge. The next section studies time and memory complexity of the methods and explains the obtained results in a more general context.

\subsection{Computational cost}
\label{sec:computational-cost-msa-pem-main}

As observed in Fig.~\ref{fig:computational_time_ex4}(b), MSA-PEM has an average running time per iteration that grows linearly with the number of stages. Since the problem dimension and the factorization cost per iteration remain constant, it is the time complexity of computing the cost function $V$ (and its derivatives) that accounts for the linear growth. In Appendix~\ref{sec:comp-cost-msa}, we discuss in detail the computation of $V$ and arrive at the time complexity of $\mathcal{O}(N K (n_y+n_u))$.

Multiple shooting estimation, on the other hand, has an average running time per iteration that decreases with the maximum simulation length $\Delta {\rm m}_{\max}$. This is displayed in Fig.\,\ref{fig:computational_time_ex4}(a). In this case, the cost of computing $V$ is $\mathcal{O}(N(n_y+n_u))$ and does not depend on the simulation length. The increase in the problem dimension, however, yields more costly factorizations (cf. Appendix~\ref{sec:comp-deriv}), which results in the increased computational cost per iteration.

For multiple shooting, shorter simulation lengths $\Delta {\rm m}_{\max}$ allow the solver to converge with fewer function evaluations. This is in agreement with the theory presented in this paper, which suggests longer simulation lengths might result in poor smoothness properties and make the optimization problem harder to solve. Applying the same reasoning to MSA-PEM, it is natural to expect that smaller values of $K$ would yield convergence with fewer iterations. This, however, does not seem to be the case for the slow time-response setting, maybe because MSA-PEM often converges to the wrong parameters in this setting (as shown in Fig.~\ref{fig:histogram_ex4}(i)).
  
In general, the number of function evaluations for MSA-PEM is lower than for multiple shooting because the resulting problem is unconstrained, rather than constrained, for which more efficient procedures are available.

Finally, MSA-PEM and multiple shooting are both amenable to parallelization, even though this is not explored in the examples. During MSA-PEM implementation, it is possible to instantiate $N$ different processes (or threads) for computing each prediction $\hat{y}[k]$ (and the corresponding derivatives) in parallel. Multiple shooting, on the other hand, subdivides the problem into $(N/\Delta {\rm m}_{\max})$ independent subproblems.

\subsection{Cost function smoothness and estimation tradeoffs}
\label{sec:tradeoffs-msapem}

\begin{table}[t]
  \caption{\textbf{Tradeoffs: multiple shooting \textit{vs} MSA-PEM.} $N_\theta$ is the number of parameters. $N_x$ is the number of states that need to be propagated through the simulation ($N_x=n_y=2$). $\Delta {\rm m}_{\max}$ is the maximum simulation length used for multiple shooting and $K$ is the number of steps ahead predicted in the MSA-PEM.}
  \centering
  \begin{tabular}{p{2.7cm}p{2.2cm}p{2.6cm}}
    \toprule
    \rowcolor[gray]{0.9}
    & multiple shoot. & MSA-PEM\\
    \midrule
    Problem dimension & $N_{\theta} + \frac{N}{\Delta {\rm m}_{\max}} \cdot N_x$& $N_{\theta}$\\
    \rowcolor[gray]{0.9}
    \# of constraints & $\frac{N}{\Delta {\rm m}_{\max}} \cdot N_x -1$& $\varnothing$\\
    Time-complexity (computing $V$) & $\mathcal{O}(N(n_y + n_u))$ & $\mathcal{O}(N K( n_y + n_u))$\\
    \rowcolor[gray]{0.9}
    Asymp. statistical properties & Independent of $\Delta {\rm m}_{\max}$ & Approach expected properties as $K \rightarrow N$\\
    \# of independent subproblems  & $\frac{N}{\Delta {\rm m}_{\max}}$ & $N$\\
    \bottomrule
  \end{tabular}
  \label{tab:tradeoffs-msa-ms}
\end{table}

Both MSA-PEM and multiple shooting allow the user to control the maximum simulation length, hence the smoothness of the cost function (cf. Section~\ref{sec:smoothness-properties}), by setting $K$ for the MSA-PEM and $\Delta {\rm m}_{\max}$ for the multiple shooting. These parameters, however, affect diverse aspects of the estimation problem, which are summarized in Table~\ref{tab:tradeoffs-msa-ms}.  Multiple shooting has an exact equivalence with the original single shooting problem regardless of the simulation length and the parameter $\Delta {\rm m}_{\max}$ offers a tradeoff between the problem dimension and smoothness properties of the cost function. MSA-PEM keeps the problem dimension fixed, and, as $K$ increases, it is possible to approach the behavior of methods with distinct statistical properties (not always in a smooth way, as shown in Fig.~\ref{fig:histogram_ex4}(i)), at the expense of an increased computational cost and worse smoothness properties.

Some refinements are proposed in the literature. For instance, \cite{farina_simulation_2011} propose to average over different values of $K$, to obtain a slower but smoother convergence to output error estimation as $K \rightarrow N$. In~\cite{farina_identification_2012}, on the other hand, the method starts from the one-step-ahead prediction solution and increases $K$ by one in each iteration until convergence. We give an numerical example using such approach in Appendix~\ref{sec:additional-experiments}.

\section{Conclusion}
\label{sec:conclusion}

The relevance of this paper lies in the rather general setting for which the proposed methods and results hold. The main technical contribution is to show that for dynamic prediction models that are non-contractive (i.e. do not converge asymptotically to a single stable point) in the region of interest, the upper bound for the Lipschitz constant and the $\beta$-smoothness blows up exponentially with the simulation length, and this can make the optimization problem very hard to solve. This was illustrated using numerical examples with systems that are non-contractive due to the presence of chaotic regions and unstable equilibrium points. Because of these regimes, the objective function becomes very intricate in some regions of the parameter space and the optimization algorithm fails to find a good solution. Even for problems that are contractive in the region of interest, multiple shooting might help in preventing the solver from getting stuck in undesirable regions of the parameter space, hence facilitating the convergence to a good solution (cf. Section~\ref{sec:pendulum-and-inverted-pendulum}).

Multiple shooting makes the simulation length a design parameter and hence allows solving optimization problems that would be infeasible in a single shooting setting. The price paid compared to single shooting methods is that a nonlinear constrained optimization problem must be solved instead of an unconstrained one. It also makes it harder to generalize to situations other than batch training, such as online training. MSA-PEM is another approach that allows some control over the smoothness of the cost function, but with different tradeoffs (see Table~\ref{tab:tradeoffs-msa-ms}) between smoothness, asymptotical properties and computational cost to be taken into consideration.

\begin{appendices}


\section{Computing the derivatives}
\label{sec:comp-deriv}

\subsection{Sensitivity equations}
\label{sec:sensitivity-eq}

Let the Jacobian matrices of $\mathbf{h}(\mathbf{x}, \underline{\mathbf{z}}; \boldsymbol{\theta})$  with respect to $\mathbf{x}$ and to $\boldsymbol{\theta}$ evaluated at the point $(\mathbf{x}[k], \underline{\mathbf{z}}[k]; \boldsymbol{\theta})$ be denoted, respectively, as $A_k$ and $B_k$. Similarly, the Jacobian matrices of $\mathbf{g}(\mathbf{x}, \underline{\mathbf{z}}; \boldsymbol{\theta})$ are denoted as $C_k$ and $F_k$. Also, we denote the Jacobian matrices of $\hat{\mathbf{y}}[k]$ with respect to $\boldsymbol{\theta}$ and to $\mathbf{x}_0$ as $J_{\boldsymbol{\theta}}[k]$ and $J_{\mathbf{x}_0}[k]$. And the Jacobian matrices of $\mathbf{x}[k]$ are denoted as $D_{\boldsymbol{\theta}}[k]$ and $D_{\mathbf{x}_0}[k]$.

A direct application of the chain rule to~(\ref{eq:nlss}) gives a recursive formula for computing the derivatives of the predicted output in relation to the parameters in the interval ${1 \le k  \le N}$:
\begin{eqnarray}
    \nonumber
  D_{\boldsymbol{\theta}}[k]
  &=& A_k D_{\boldsymbol{\theta}}[k-1] +B_k\text{ for }D_{\boldsymbol{\theta}}[0] = \mathbf{0}\\
  \label{eq:sensitivity_eq_params}
  J_{\boldsymbol{\theta}}[k] &=& C_k D_{\boldsymbol{\theta}}[k] + F_k.
\end{eqnarray}
A similar recursive formula may be used for computing the derivatives of the predicted output in relation to the initial conditions:
\begin{eqnarray}
  \label{eq:sensitivity_eq_states}
  D_{\mathbf{x}_0}[k]
  &=& A_k D_{\mathbf{x}_0}[k-1] \text{ for }D_{\mathbf{x}_0}[0] = \mathbf{I} \\\nonumber
  J_{\mathbf{x}_0}[k]  &=& C_k D_{\mathbf{x}_0}[k].
\end{eqnarray}

Finally,  we define $D[k] = [D_{\boldsymbol{\theta}}[k], D_{\mathbf{x}_0}[k]]$ and \\ $J[k] = [J_{\boldsymbol{\theta}}[k], J_{\mathbf{x}_0}[k]]$.

\subsection{Single shooting}

For the cost function $V$ defined as in~(\ref{eq:cost_function}), its gradient $\nabla V$ is given by:
\begin{equation}
  \label{eq:grad_objective_fun}
  \nabla V = \tfrac{2}{N}\sum_{k=1}^{N}  J[k]^T (\hat{\mathbf{y}}[k] -\mathbf{y}[k]),
\end{equation}
Its Hessian $\nabla^2 V$ is given by:
\begin{equation}
  \label{eq:hess_objective_fun}
  \nabla^2 V = \tfrac{2}{N}\sum_{k=1}^{N} \left(J[k]^T  J[k]  + \mathbf{S}[k]\right).
\end{equation}
where $\mathbf{S}[k] = \sum_{j=1}^{N_y}\hat{y}_j[k] \nabla^2 \hat{y}_j[k]$. Ignoring $\mathbf{S}[k]$ is a common approximation used in least-squares algorithms, that will also be used here when computing derivatives numerically.

\subsection{Multiple shooting}

In order to solve the problem using the sequential quadratic programming solver~\cite{lalee_implementation_1998}
we must be able to compute:
i) The cost function $V^M$; ii) its gradient $\nabla V^M$; iii) the constraints;
iv) the Jacobian matrix of the constraints (which can be represented using a sparse
representation); and, v) for any given vector $\mathbf{p}$, the product of the Lagrangian\footnotemark~Hessian
and the vector $\mathbf{p}$ (the full Lagrangian Hessian matrix does not need to be computed). The following sequence provides a way of computing all derivatives required by the optimizer.

\footnotetext{The Lagrangian is given by:
  $\mathcal{L}(\boldsymbol{\phi}, \boldsymbol{\lambda}) =
  V(\boldsymbol{\phi}) +  \boldsymbol{\lambda}^T\mathbf{c}(\boldsymbol{\phi})$.}

\begin{alg}[Derivatives]    
    For a given parameter $\boldsymbol{\theta}$ and set
    of initial conditions:
    \begin{enumerate}
    \item For $i = 1, \cdots, M$, do:
      \begin{enumerate}
      \item For $ k = \text{m}_{i}+1, \cdots, \text{m}_{i+1}$:
          \begin{enumerate}
          \item
          Compute $\mathbf{x}^i[k]$ and $\hat{\mathbf{y}}^i[k]$  with~(\ref{eq:nlss}).
    
          \item Compute $A_k$, $B_k$, $C_k$ and $F_k$.
          \item Compute $D^i[k]$ and $J^i[k]$ with the sensitivity equations.
        \end{enumerate}
      \item Compute $V_i$ using
        Eq.~(\ref{eq:Vi}).
      \item Compute $\nabla V_i$ using a formula equivalent to~(\ref{eq:grad_objective_fun}).
      \item Approximate the product of the Hessian
        with a given vector,
        $\nabla^2 V_i \mathbf{p}$, using
        the first terms from a expression equivalent to~(\ref{eq:hess_objective_fun})
      \end{enumerate}
    \item Compute $V^M$ with~(\ref{eq:cost_function_ms});
    \item Compute $\nabla V^{M} = \sum_{i=1}^M \tfrac{\Delta {\rm m}_i}{N} ~ \nabla V_{i}$;
    \item Compute the value of the constraint from the values of $\mathbf{x}^i[\text{m}_{i+1}]$, $i = 1, \cdots, M$;
    \item Compute the Jacobian matrix of the constraints from
      $J^i[\text{m}_{i+1}]$, $i = 1, \cdots, M$;
    \item Compute
      $\nabla^2 V^M \mathbf{p} = \sum_{i=1}^M \tfrac{\Delta {\rm m}_i}{N} \nabla^2 V_{i}\mathbf{p}$;
    \item Compute the product of the Hessian $\boldsymbol{\lambda}^T\mathbf{c}(\boldsymbol{\phi})$
      with a vector $\mathbf{p}$ using 2-point finite differences;
    \item  Compute the product of the Lagrangian Hessian  and a vector
      $\nabla^2 \mathcal{L}(\boldsymbol{\phi}, \boldsymbol{\lambda})\mathbf{p}$,
      summing the Hessians computed in steps 6 and 7.
    \end{enumerate}
\end{alg}

Some approximations were used for computing the second derivatives:
1) for computing the Hessian of the objective function, the standard least-squares
approximation for the Hessian is used; and, 2) for computing the Hessian
of the constraint we use finite-difference approximation. The use of finite differences
here comes inexpensively because we only need to evaluate the Hessian times a vector,
and not the full matrix. Hence, it can be done at the cost of an extra
Jacobian matrix evaluation.

Notice that step (1) from the above algorithm can be parallelized, with different processes
(or threads) performing the computation for different values of $i$.

\section{Proofs}
\label{sec:proof_thm_ms_lipschitz}

\subsection{Preliminary results}

\begin{lem}
\label{thm:lipschtz_sum_prod}
Let $\mathbf{f}$ and $\mathbf{g}$ be two Lipschitz functions on~$\Omega$ with  Lipschitz constants $L_f$ and $L_g$. Then,
\begin{enumerate}[a)]
    \item $\mathbf{f}+\mathbf{g}$ is also a Lipschitz function on~$\Omega$ with the (best) Lipschitz constant upper bounded by $(L_f+L_g)$;
    \item if, additionally, $\mathbf{f}$ and $\mathbf{g}$ are bounded by $M_f$ and $M_g$ on~$\Omega$, then  $\mathbf{f}\mathbf{g}$ is also a Lipschitz function on~$\Omega$ with  the (best) Lipschitz constant upper bounded by $(L_f M_g+L_g M_f)$.
\end{enumerate}
\end{lem}

\subsection{Proof of Theorem \ref{thm:lipschitz} (a)}

Assume two different trajectories resulting from simulating the system~(\ref{eq:nlss}) with parameters and initial conditions $(\mathbf{x}_0, \boldsymbol{\theta})$ and $(\mathbf{w}_0, \boldsymbol{\phi})$, respectively. We denote the corresponding trajectories by $\mathbf{x}$ and $\mathbf{w}$. Let us call:
\begin{equation}
    \|\Delta \hat{\mathbf{y}}[k]\| = \| \textbf{g}(\mathbf{x}[k], \underline{\mathbf{z}}[k]; \boldsymbol{\theta}) - \textbf{g}(\mathbf{w}[k], \underline{\mathbf{z}}[k]; \boldsymbol{\phi})\| .
\end{equation}

\noindent
Because $\textbf{h}$ and $\textbf{g}$ are Lipschitz in ${(\mathbf{x}, \boldsymbol{\theta})}$ we have:
\begin{eqnarray*}
\|\textbf{h}(\mathbf{x}, \underline{\mathbf{z}}, \boldsymbol{\theta}) - \textbf{h}(\mathbf{w},  \underline{\mathbf{z}}, \boldsymbol{\phi})\|^2 \le L_h^2 \left(\|\mathbf{x} - \mathbf{w}\|^2 + \|\boldsymbol{\theta} - \boldsymbol{\phi}\|^2 \right), \\
\|\textbf{g}(\mathbf{x}, \underline{\mathbf{z}}, \boldsymbol{\theta}) - \textbf{g}(\mathbf{w}, \underline{\mathbf{z}}, \boldsymbol{\phi})\|^2 \le L_g^2 \left(\|\mathbf{x} - \mathbf{w}\|^2 + \|\boldsymbol{\theta} - \boldsymbol{\phi}\|^2 \right),
\end{eqnarray*}
for all $(\mathbf{x}, \underline{\mathbf{z}}, \boldsymbol{\theta})$ and $(\mathbf{w}, \underline{\mathbf{z}}, \boldsymbol{\phi})$ in $(\Omega_{\mathbf{x}},\Omega_{\underline{\mathbf{z}}}, \Omega_{\boldsymbol{\theta}})$. Applying these relations recursively we get that:
\begin{equation*}
    \|\Delta \hat{\mathbf{y}}[k]\|^2 \le L_g^2 L_h^{2k} \|\boldsymbol{x}_0 - \boldsymbol{w}_0 \|^2 + L_g^2 \left(\sum_{\ell=0}^k L_h^{2\ell}\right)\|\boldsymbol{\theta} - \boldsymbol{\phi}\|^2.
\end{equation*}

\noindent
Since $L_h$ is positive,  the constant multiplying the second term in the above equation is always larger than the constant multiplying the first one. Hence, taking the square root on both sides of the above inequality and after simple manipulations, we get:
\begin{equation}
    \label{eq:ineq_dk}
    \|\Delta \hat{\mathbf{y}}[k]\| \le  L_gS(k)\|[\boldsymbol{\theta}, \boldsymbol{x}_0]^T - [\boldsymbol{\phi}, \boldsymbol{w}_0]^T \|.
\end{equation}
where:
\begin{equation}
    \label{eq:Sk}
    S(k) = \sqrt{\sum_{\ell=0}^k L_h^{2\ell}} =     
    \begin{dcases}
    \sqrt{k + 1} & \text{ if }L_h = 1 \\
    \sqrt{\frac{L_h^{2k+2} - 1}{L_h^2 - 1}} & \text{ if }L_h \not= 1 .
    \end{dcases}
\end{equation}

Since $\Omega$ is compact and $\hat{\mathbf{y}}[k]$ is a (Lipschitz) continuous function of the parameters and initial conditions, then $\hat{\mathbf{y}}[k]$ is bounded in $\Omega$, i.e. $\|\hat{\mathbf{y}}[k]\| \le M(k)$.  And, it follows from~(\ref{eq:ineq_dk}) and from the existence of an invariant set\footnote{There are multiple ways to guarantee the invariant set premise will hold, but a very simple way is to just choose $\textbf{h}$ such that $\textbf{h}(\mathbf{0}, \underline{\mathbf{z}}; \mathbf{0}) = \textbf{0}$. In this case, $\{\mathbf{0}\}$ is an invariant set and if $\Omega_{\boldsymbol{\theta}}$ contain this point the premise is satisfied. For this specific case, one can just choose $[\boldsymbol{\phi}, \boldsymbol{w}_0] = \mathbf{0}$ and it follows from~(\ref{eq:ineq_dk}) that $\|\hat{\mathbf{y}}[k]\| \le  L_gS(k)\|[\boldsymbol{\theta}, \boldsymbol{x}_0]\| = \mathcal{O}(S(k))$. The more general case, for any invariant set, follows from a similar deduction.} in $\Omega$ that $M(k) = \mathcal{O}(S(k))$.

The following inequality follows from~(\ref{eq:cost_function}):
\begin{equation}
    \label{eq:ineq_V1}
    |V(\boldsymbol{\theta}, \boldsymbol{x}_0) - V(\boldsymbol{\phi}, \boldsymbol{w}_0)| \le  \tfrac{2}{N}\sum_{k=1}^{N}  (L_y + M(k)) \|\Delta \hat{\mathbf{y}}[k]\| ,
\end{equation}

\noindent
where ${L_y = \max_{1 \leq k \leq N} \|\mathbf{y}[k]\|}$. And, by putting together~(\ref{eq:ineq_V1}) and~(\ref{eq:ineq_dk}):
\begin{equation*}
|V(\boldsymbol{\theta}, \boldsymbol{x}_0) - V(\boldsymbol{\phi}, \boldsymbol{w}_0)| \le \\
L_{V_1} \left\|[\boldsymbol{x}_0, \boldsymbol{\theta}]^T - [\boldsymbol{w}_0, \boldsymbol{\phi}]^T \right\|,
\end{equation*}

\noindent
for ${L_{V} = \left(\tfrac{2 L_g}{N}\sum_{k=1}^{N} (L_y + M(k))S(k)\right)}$. The asymptotic analysis of this expression with regard to $N$ yields~(\ref{eq:asymptotic_L}).

\subsection{Proof of Theorem \ref{thm:lipschitz} (b)}

It follows from~(\ref{eq:grad_objective_fun}) that:
\vspace*{-2.5\baselineskip}
\begin{small}
\begin{equation}
    \label{eq:ineq_grad_V}
    \|\nabla V(\boldsymbol{\theta}, \boldsymbol{x}_0) - \nabla V(\boldsymbol{\phi}, \boldsymbol{w}_0)\|  \le      \tfrac{2}{N}\sum_{k=1}^{N}  L_y \|\Delta J[k]\|
    + \|\Delta (J[k] \hat{\mathbf{y}}[k])\| ,
\end{equation}
\end{small}

\noindent
where we have used the notation $\Delta J[k]$  to denote the difference between $J[k]$ evaluated at $(\boldsymbol{\theta}, \boldsymbol{x}_0)$ and $(\boldsymbol{\phi}, \boldsymbol{w}_0)$. Analogously, $\Delta (J[k] \hat{\mathbf{y}}[k])$ denote the difference between $J[k]\hat{\mathbf{y}}[k]$ evaluated at the two distinct points.

From equation (\ref{eq:sensitivity_eq_params}) and (\ref{eq:sensitivity_eq_states}) it follows that:
\vspace*{-2\baselineskip}
\begin{small}
\begin{equation}
    \label{eq:close_formula}
    J_{\boldsymbol{\theta}}[k] = C_k \sum_{\ell=1}^{k} \left(\prod_{j=1}^{k-\ell} A_{k-j+1}\right) B_\ell  + F_k;~~
    J_{\mathbf{x}_0}[k] = C_k \prod_{\ell=1}^{k} A_{k-\ell+1}.
\end{equation}
\end{small}

\noindent
Since, the Jacobian of $\textbf{h}$ is Lipschitz with Lipschitz constant $L_h'$, it follows that:
\begin{eqnarray}
\|\Delta A_j\|^2 \le (L_h')^2\left(\|\mathbf{x}[j] - \mathbf{w}[j]\|^2 + \|\boldsymbol{\theta} - \boldsymbol{\phi}\|^2\right).
\end{eqnarray}
Using a procedure analogous to the one used to get Eq.~(\ref{eq:ineq_dk}), it follows that:
\begin{eqnarray}
\label{eq:ineq_aj}
\|\Delta A_j\| \le L_h' S(j)~~\|[\boldsymbol{\theta}, \boldsymbol{x}_0]^T - [\boldsymbol{\phi}, \boldsymbol{w}_0]^T \|,
\end{eqnarray}
where $S(j)$ is defined as in~(\ref{eq:Sk}). An identical formula holds for $B_j$ and a similar formula, replacing $L_h'$ with $L_g'$, holds for $C_j$ and $F_j$. 

Since $\textbf{h}$ and $\textbf{g}$ are Lipschitz with Lipschitz constants $L_h$ and $L_g$ it follows that $\|A_j\| \le L_h$, $\|B_j\| \le L_h$, $\|C_j\| \le  L_g$ and $\|F_j\| \le L_g$. Hence, it follows from~(\ref{eq:ineq_dk}),~(\ref{eq:close_formula}), (\ref{eq:ineq_aj}) and the repetitive application of Lemma~\ref{thm:lipschtz_sum_prod} that $\|\Delta J_{\boldsymbol{\theta}}[k]\|$, $\|\Delta J_{\mathbf{x}_0}[k]\|$, $\|\Delta (J_{\boldsymbol{\theta}}[k]\hat{\mathbf{y}}[k])\|$ and $\|\Delta (J_{\mathbf{x}_0}[k]\hat{\mathbf{y}}[k])\|$
are upper bounded by $\|[\boldsymbol{\theta}, \boldsymbol{x}_0]^T - [\boldsymbol{\phi}, \boldsymbol{w}_0]^T \|$ multiplied by the following constants:
\begin{eqnarray*}
\nonumber
L_{J_{\boldsymbol{\theta}}}(k) =\sum_{\ell=1}^k P(k, \ell) + L_g' S(k)&;~& L_{J_{\mathbf{x}_0}}(k) = P(k, 1)\\
L_{J_{\boldsymbol{\theta}}\hat{\mathbf{y}}}(k) = \sum_{\ell=1}^k Q(k, \ell)  +  T(k)S(k)&;~&
L_{J_{\mathbf{x}_0}\hat{\mathbf{y}}}(k) = Q(k, 1)  ,
\end{eqnarray*}

\noindent
where $T(k) = (L_g' M(k) + L_g^2)$ and:
\begin{eqnarray*}
    P(k, \ell) &=& L_h^{k-\ell}\left(L_gL_h'\sum_{j=\ell}^k S(j) +  L_hL_g'S(k)\right) \\
     Q(k, \ell) &=& L_h^{k-\ell}\left(M(k)L_gL_h'\sum_{j=\ell}^k S(j) +  L_hT(k)S(k)\right).
\end{eqnarray*}

\noindent
Hence,
\begin{equation*}
    \|\nabla V(\boldsymbol{\theta}, \boldsymbol{x}_0) - \nabla V(\boldsymbol{\phi}, \boldsymbol{w}_0)\|  \le      L_{V}' \|[\boldsymbol{\theta}, \boldsymbol{x}_0]^T - [\boldsymbol{\phi}, \boldsymbol{w}_0]^T \|,
\end{equation*}

\noindent
where 
\begin{scriptsize}
\begin{eqnarray*}
    L_{V}' = \tfrac{2}{N}\sum_{k=1}^{N} \left(L_y(L_{J_{\boldsymbol{\theta}}}(k) + L_{J_{\mathbf{x}_0}}(k) ) + L_{J_{\boldsymbol{\theta}}\hat{\mathbf{y}}}(k) + L_{J_{\boldsymbol{\theta}}\hat{\mathbf{y}}}(k)\right) .
\end{eqnarray*}
\end{scriptsize}
Putting everything together the asymptotic analysis of $L_{V}'$ results in~(\ref{eq:asymptotic_L'}).

\section{Lipschitz analysis for the multiple shooting}
\label{sec:lipsch-analys-mult}

The next theorem follows from basic inequality manipulation and relates the Lipschitzness and $\beta$-smoothness of the cost function $V^M$ with that of its components $V_i$.

\begin{thm}
\label{thm:ms_cost}
Defining $V^M$ as in~(\ref{eq:cost_function_ms}), if each component $V_i$ is Lipschitz continuous with constant $L_{V_i}$ then $V^M$ is also Lipschitz with constant equal to or smaller than $L_{V^M} =\max(L_{V_1}, \cdots, L_{V_M})$. Additionally, if the gradient of each component $\nabla V_i$ is Lipschitz continuous with constant $L_{V_i}'$ then $\nabla V^M$  is also Lipschitz with constant equal to or smaller than $L_{V^M}' = \max(L_{V_1}', \cdots, L_{V_M}')$.
\end{thm}

\begin{pf}
For $\boldsymbol{\theta}_{\text{ext}} = (\boldsymbol{\theta}, \mathbf{x}_{0}^1, \cdots, \mathbf{x}_{0}^M)$ and $ \boldsymbol{\phi}_{\text{ext}} = (\boldsymbol{\phi}, \mathbf{w}_{0}^1, \cdots, \mathbf{w}_{0}^M)$ we have that:
\begin{multline*}
|V^M(\boldsymbol{\theta}_{\text{ext}} ) - V^M(\boldsymbol{\phi}_{\text{ext}} )| \le \\
\sum_{i=1}^M \tfrac{\Delta {\rm m}_i}{N} ~ |V_{i}(\boldsymbol{\theta}, \mathbf{x}_{0}^i) -V_i(\boldsymbol{\phi}, \boldsymbol{w}_0^i)| \le \\
\sum_{i=1}^M \tfrac{\Delta {\rm m}_i}{N} L_{V_i} \|[\boldsymbol{\theta}, \boldsymbol{x}_0^i]^T - [\boldsymbol{\phi}, \boldsymbol{w}_0^i]^T \| \le \\
L_{V^M} \|\boldsymbol{\theta}_{\text{ext}} - \boldsymbol{\phi}_{\text{ext}}\|,
\end{multline*}

\nonumber
where $L_{V^M} = \max(L_{V_1}, \cdots, L_{V_M})$. And similarly, ${L_{V^M}' = \max(L_{V_1}', \cdots, L_{V_M}')}$, which yields the second result.
\end{pf}

\section{Asymptotic properties of prediction error methods}
\label{sec:asymptotic-properties}

\subsection{Notation and data generation process}
Consider $\mathbf{y}[k]$ and $\mathbf{u}[k]$ to be one realization of the random variables $\mathbf{Y}[k]$ and $\mathbf{U}[k]$. And denote:
\begin{eqnarray*}
    \underline{\mathbf{U}}[k] &=& [\mathbf{U}[k], \cdots, \mathbf{U}[k-n_u]], \\
    \underline{\mathbf{Y}}[k-1] &=& [\mathbf{Y}[k-1],  \cdots, \mathbf{Y}[k-n_y]].
\end{eqnarray*}
Hence, rewriting the data generation difference equations for nonlinear ARX, output error and ARMAX (see Sections~\ref{sec:narx},~\ref{sec:noe} and~\ref{sec:narmax}) with the random variables yield, respectively:
\vspace*{-0.5\baselineskip}
\begin{scriptsize}
\begin{eqnarray*}
  \label{eq:narx_data_model}
  \bullet &&\mathbf{Y}[k] = \mathbf{f}^*(\underline{\mathbf{Y}}[k-1],
  \underline{\mathbf{U}}[k]) + \mathbf{V}[k]\\
  \bullet &&\left\{\begin{aligned}
      \bar{\mathbf{Y}}[k] &= \mathbf{f}^*(\bar{\mathbf{Y}}[k-1], \hdots, \bar{\mathbf{Y}}[k-n_y],
      \underline{\mathbf{U}}[k])\\
      \mathbf{Y}[k] &= \bar{\mathbf{Y}}[k] + \mathbf{V}[k],
    \end{aligned}\right. \\
  \bullet&&\mathbf{Y}[k] = \mathbf{f}^*(\underline{\mathbf{Y}}[k-1],
  \underline{\mathbf{U}}[k],
    \mathbf{V}[k-1], \hdots,
                    \mathbf{V}[k-n_v]) + \mathbf{V}[k].
\end{eqnarray*}
\end{scriptsize}
Here $\mathbf{V}[k]$ is a random variable representing the noise that is injected in the system. Notice that there is a deterministic additive relation between $\mathbf{Y}[k]$ and $\mathbf{V}[k]$. Hence, if $\mathbf{Y}[k]$ and $\mathbf{U}[k]$ are determined, so is $\mathbf{V}[k]$, or, conversely, if $\mathbf{V}[k]$ and $\mathbf{U}[k]$ are determined, so is $\mathbf{Y}[k]$.

\subsection{Optimal output prediction}

Let us define the \textit{optimal output prediction} at time $k$ as the following conditional expectation:
\begin{equation}
  \label{eq:conditional_expectation}
  \hat{\mathbf{y}}_*[k] = E\left\{\mathbf{Y}[k]~\Big\vert~ \underline{\mathbf{U}}[k] = \underline{\mathbf{u}}[k],~ \underline{\mathbf{Y}}[k-1] = \underline{\mathbf{y}}[k-1] \right\},
\end{equation}
\noindent
which is, in the least square sense, the best prediction for the output
given its previous values.\footnote{
This prediction provides the smallest squared conditional expected error
between the predicted and observed values: $
  \hat{\mathbf{y}}_*[k] = \arg_{\hat{\mathbf{Y}}} \min
               E\left\{\|\mathbf{Y}[k] - \hat{\mathbf{Y}}\|^2 ~\Big\vert~\underline{\mathbf{u}}[k], \underline{\mathbf{y}}[k-1] \right\}.$
}

For the nonlinear ARX, output error and ARMAX models the \textit{optimal output predictions} are given, respectively, by:
\vspace*{-2.5\baselineskip}
\begin{small}
\begin{eqnarray*}
  \label{eq:narx_optimal_predictor}
  \bullet&&\hat{\mathbf{y}}_*[k] = \mathbf{f}^*(\underline{\mathbf{y}}[k-1],
        \underline{\mathbf{u}}[k]).\\
  \label{eq:noe_optimal_predictor}
  \bullet&&\left\{\begin{aligned}
      \bar{\mathbf{y}}[k] &= \mathbf{f}^*(\bar{\mathbf{y}}[k-1], \hdots, \bar{\mathbf{y}}[k-n_y],
      \underline{\mathbf{u}}[k])\\
      \hat{\mathbf{y}}_*[k] &= \bar{\mathbf{y}}[k], \label{eq:true_system}
    \end{aligned}\right.\\
  \label{eq:narmax_optimal_predictor}
  \bullet&&\left\{\begin{aligned}
        &\tilde{\mathbf{v}}[k] = \mathbf{y}[k] - \mathbf{f}^*(\underline{\mathbf{y}}[k-1],
        \underline{\mathbf{u}}[k],
       \tilde{\mathbf{v}}[k-1], \hdots,
        \tilde{\mathbf{v}}[k-n_v])\\
         &\hat{\mathbf{y}}_*[k] = \mathbf{f}^*(\underline{\mathbf{y}}[k-1],
                          \underline{\mathbf{u}}[k],
                          \tilde{\mathbf{v}}[k-1], \hdots,
                          \tilde{\mathbf{v}}[k-n_v])
                        \end{aligned}\right.
\end{eqnarray*}
\end{small}
which follows from a direct application of the definition~(\ref{eq:conditional_expectation}) to the stochastic difference equations that are assumed for the data generation in each case. 

\subsection{Ideal cost function}
Ideally, the model predicted output $\hat{\mathbf{y}}[k]$ should be
as close as possible to the optimal one $\hat{\mathbf{y}}_*[k]$.
The distance between the two series is quantified by the cost function:
\begin{equation}
  \label{eq:optimal_cost_function}
  V_* = \frac{1}{N}\sum_{k=1}^N \|\hat{\mathbf{y}}_*[k] - \hat{\mathbf{y}}[k]\|^2.
\end{equation}
Now, back to the examples, given a parametrized function $\mathbf{f}_{\boldsymbol{\theta}}$ with the correct model structure (that is: there exist $\boldsymbol{\theta}^*$  such that $\mathbf{f}_{\boldsymbol{\theta}^*} = \mathbf{f}^*$), it follows from Eqs.~(\ref{eq:narx_optimal}),~(\ref{eq:noe_optimal}) and~(\ref{eq:narmax_optimal}) that $\boldsymbol{\theta}^*$ yields  $V_* = 0$, and hence $\boldsymbol{\theta}^*$ is a minimizer of $V_*$.

\subsection{Uniform convergence of $V$}
The optimal cost function $V_*$ is not available for optimization. Nevertheless, under some mild regularity conditions, it has been proved that $V \rightarrow V_*$  in probability as  $N \rightarrow \infty$ and that \textit{this convergence is uniform}~\cite{ljung_convergence_1978}.

If $V_*$ has a single minimum $\boldsymbol{\theta}^*$ and $V$ is convex in a convex set containing $\boldsymbol{\theta}^*$ the minimizer of $V$ converges to the minimizer of $V_*$~\cite[Theorem 2.1]{newey_large_1994a}.  Alternative conditions for this to hold are given in \cite{ljung_convergence_1978}.  For our three examples, this would imply $\mathbf{f}_{\theta}\rightarrow \mathbf{f}^*$ (convergence in probability for $N\rightarrow \infty$). If the conditions are not satisfied, the uniform convergence, at least, guarantees that the minimizer has an equivalent performance (as $N\rightarrow \infty$). Additionally,  in~\cite{ljung_asymptotic_1980} it is shown that if the solution is unique the estimator has an  asymptotic normal distribution.

\section{Computational cost of MSA-PEM}
\label{sec:comp-cost-msa}

As mentioned in Section~\ref{sec:computational-cost-msa-pem-main}, the computation of the cost function $V$ (and its derivatives) for the MSA-PEM method has linear time complexity with $K$. The rate at which the time complexity grows, however, depends on implementation choices. We use the model estimation described in the experiment in Section~\ref{sec:exampl-estim-outp} to explain the tradeoffs of those choices. Propagating $K=3$ steps-ahead the linear system $y[k] = \theta_1 y[k] + \theta_2 y[k-2] + \theta_3 u[k-1]$ can be performed in two ways. The first is to introduce intermediate variables:
\begin{align*}
      \tilde{y}_k[k-2]&\leftarrow \theta_1 y[k-3] + \theta_2 y[k-4] + \theta_3 u[k-3],\\
      \tilde{y}_k[k-1]&\leftarrow  \theta_1 \tilde{y}_k[k-2] + \theta_2 y[k-3] + \theta_3 u[k-2]\\
      \hat{y}[k]&\leftarrow  \theta_1 \tilde{y}_k[k-1] + \theta_2\tilde{y}_k[k-2] + \theta_3 u[k-1].
\end{align*}
The other is to simplify the computation in advance, resulting in a system of higher  order:
\begin{multline*}
  \hat{y}[k]\leftarrow  \theta_1'y[k-3] + \theta_2'y[k-4]\\
    + \theta_3' u[k-1]+ \theta_4' u[k-2] + \theta_5' u[k-3],
  \end{multline*}
  where the $\boldsymbol{\theta}'$ coefficients depend on $\boldsymbol{\theta}$ (for instance  $\theta_1' = \theta_1^3-2\theta_1\theta_2$) and are computed in advance. For the first approach, computing the cost function has time-complexity $\mathcal{O}(NK(n_y+n_u))$; for the second, the computational cost is  slightly smaller: $\mathcal{O}(N(K+n_u+n_y))$. The reduced computational cost is obtained  because some computations are performed in advance and stored. Hence, the additional time-efficiency comes at the cost of storing the computation in advance, and the memory of the first method is $\mathcal{O}(n_y+n_u)$, while in the second approach the memory complexity is $\mathcal{O}(K+n_y+n_u)$.

  For linear systems, it is easy to find the optimal simplified computation in advance by using Diophantine equations~\cite{farina_simulation_2011}. In the experiment in Section~\ref{sec:exampl-estim-outp}, however, we use the less time-efficient approach. The general tone of our discussion justifies this choice since this implementation can easily be applied to generic systems. And, while heuristic solutions might be employed to simplify the computations in advance, there is no general approach for all types of nonlinear systems. Regardless of this design choice, however, the observed linear growth of the computational cost with $K$ is expected.

\section{Additional experiments}
\label{sec:additional-experiments}

\subsection{Neural network for modeling pilot plant}
\begin{figure}[tpb]
\includegraphics{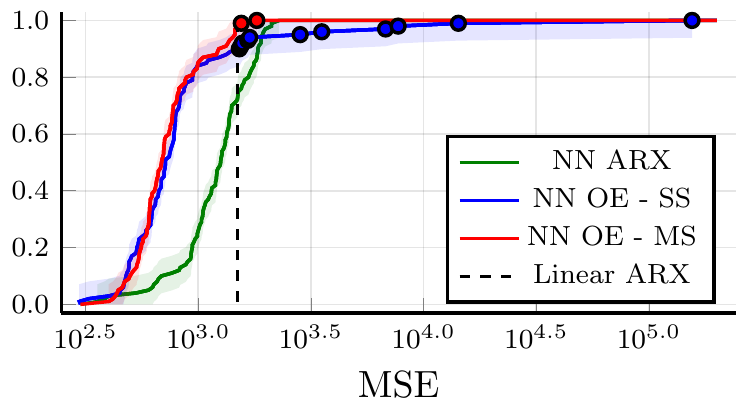}
\caption{\textbf{Neural network performance}. Displays the empirical cumulative distribution of the free-run simulation MSE over the validation dataset. The results obtained in~\cite{ribeiro_parallel_2018} for an  ARX neural network (NN ARX) and an (single shooting) output error neural network (NN OE - SS) are displayed together with the result obtained estimating the parameters using multiple shooting (NN OE - MS). A Linear ARX model is considered as a baseline and is displayed by the dashed line. The multiple shooting estimation uses $\Delta \rm{m}_{\max} = 3$ and the training is restricted to 2000 iterations of the optimization algorithm or until either the gradient or the step size drops below $10^{-12}$. The other models were estimated exactly as in~\cite{ribeiro_parallel_2018}. All the neural network models have 10 nodes in the hidden layer, ${n_y=n_u=1}$ and  were trained with the same training dataset. Each curve is the result of 100 realizations and, for each realization, the neural network initial weights $w_{i,j}^{(n)}$ are drawn from a normal distribution with zero-mean and standard deviation ${\sigma=(N_{s (n-1)})^{-0.5}}$  and the bias terms $\gamma_i^{(n)}$ are initialized with zeros~\cite{lecun_efficient_1998}.  Realizations of \texttt{NN OE - SS} and \texttt{NN OE - MS} that perform worse than the baseline are indicated respectively as blue, \protect\includegraphics{img/bluecircle.pdf}, and red circles, \protect\includegraphics{img/redcircle.pdf}. Confidence intervals (95\%) are displayed as shaded regions around the estimated cumulative distribution, these have been computed using the Dvoretzky-Kiefer-Wolfowitz inequality~\cite{dvoretzky_asymptotic_1956}.}
\label{fig:distr_nn_pilotplant}
\end{figure}
\label{sec:neural-network-pilot-plant}

This example makes use of data from the level process station described in Example~1 from~\cite{ribeiro_parallel_2018}. As in the original paper, we use a neural network to  model the water column height as a function of the voltage applied to a control valve that modulates the water flow.  We compare three different training methods: i) minimizing the one-step-ahead prediction error (\texttt{NN ARX}) ii) minimizing the free-run simulation error using single shooting method (\texttt{NN OE - SS}); and, iii) minimizing the free-run simulation error using multiple shooting method (\texttt{NN OE - MS}).

The neural network (NN) training depends on the weight initialization, hence the performance of the neural network can be regarded as a random variable and is displayed in Fig.~\ref{fig:distr_nn_pilotplant}, which compares the empirical cumulative distribution of the \textit{mean square error} (MSE) over the validation dataset for the three methods. A linear ARX model ($n_y=1$ and $n_u=1$) was trained and tested under the same conditions to serve as the baseline. Methods (i) and (ii) and the linear ARX baseline were described in~\cite{ribeiro_parallel_2018}. Method (iii) is introduced in this paper.

The cumulative distribution function gives, for each $x$-axis value, the probability of the method to yield a validation MSE smaller than or equal to this value. It was estimated from 100 realizations of the neural network training procedure. Fig.~\ref{fig:distr_nn_pilotplant} shows that for more than 90\% of the realizations, estimating the parameters by minimizing the free-run simulation (\texttt{NN OE}) offers significant advantages over the minimization of the one-step-ahead error (\texttt{NN ARX}). When using a standard single shooting formulation, however, it also makes the parameter estimation procedure more sensitive to the initial conditions, with the algorithm yielding some really bad results for some initial choices~\cite{ribeiro_parallel_2018}. This results in a long-tailed distribution for the MSE (Fig.~\ref{fig:distr_nn_pilotplant}). More precisely, in 10 out of 100 realizations the \texttt{NN OE - SS} model yields a performance that is inferior to the linear ARX baseline, some of the realizations worse than the linear baseline by a factor of 100. The performance of the \texttt{NN OE - SS} and \texttt{NN OE - MS} is very similar for 90\% of the realizations, the tail of the distribution, however, is very different, with the multiple shooting procedure rarely producing very bad results. In order to highlight the differences, results where \texttt{NN OE - SS} and \texttt{NN OE - MS} are worse than the baseline are presented, respectively, as blue and red circles in Fig.~\ref{fig:distr_nn_pilotplant}. 

This example illustrates how  the use of multiple shooting alleviates the problem of high sensitivity to initial conditions, making it possible to estimate output error models with extra robustness against variations of the initial conditions and lower probability of getting trapped at local minima with very bad performance.

\begin{figure*}[t]
  \centering
  \subfloat[][$K = 2$]{\hspace{-0.25cm}\includegraphics{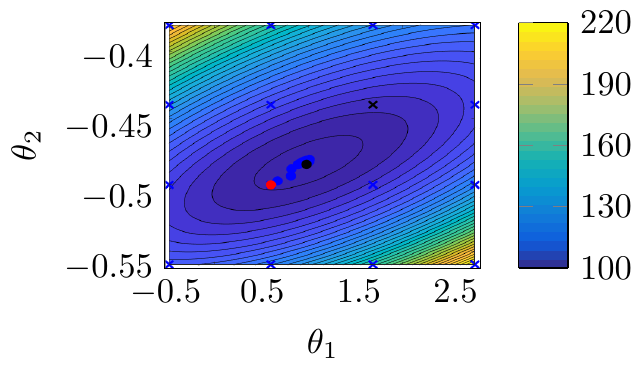}}
  \subfloat[][$K = 10$]{\hspace{-0.2cm}\includegraphics{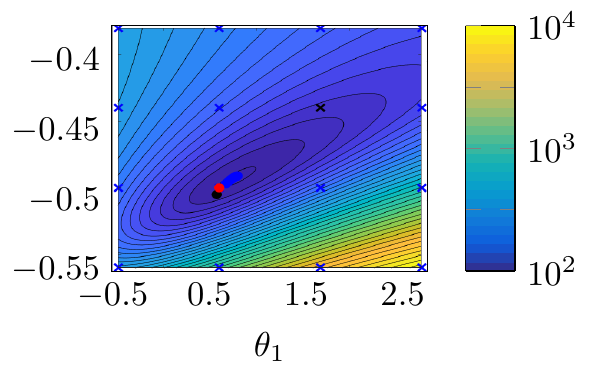}}
  \subfloat[][$K = 30$]{\hspace{-0.2cm}\includegraphics{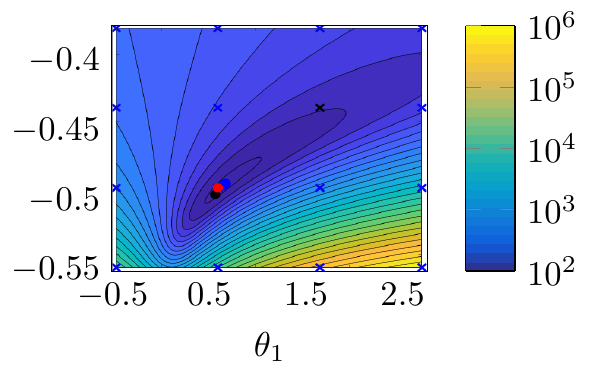}}
    \caption{\textbf{Contour plot of the MSA-PEM cost function.} The figures correspond to the $K$-step-ahead cost function for increasing values of $K$. The true parameters are indicated with a red circle \protect\includegraphics{img/redcircle.pdf}. Initial conditions are indicated with blue crosses \protect\includegraphics{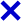}\ and the resulting estimated parameters with blue circles \protect\includegraphics{img/bluecircle.pdf}. The default method in \cite{farina_identification_2012} is to start from the one-step-ahead solution (\protect\includegraphics{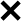}\,) and increase $K$ by one in each iteration until convergence. The intermediary results (up to the  $K = {2, 10, 30}$ iteration) for the refined implementation are displayed in the plot and are indicated with a black circle \protect\includegraphics{img/blackcircle.pdf}. In (a), the color scale is linear, and in (b) and (c), the color scale is logarithmic.}
    \label{fig:smoothness_ex5}
\end{figure*}

This example also shows the limitations of the multiple shooting formulation. The training time for~\texttt{NN OE - MS} model is 282 seconds, for \texttt{NN OE - SS} model is 3.9 seconds, and for \texttt{NN ARX} model is 3.3 seconds.  This means that the single shooting parameter estimation could be repeated, roughly, 70 times for each multiple shooting run. Hence solving the single shooting problem several times and choosing the best result would also avoid very bad solutions and could, still, be computationally less expensive than solving the multiple shooting  problem. The longer training time is due to two factors: i) per iteration the multiple shooting approach takes, roughly, 3.5 times more than the single shooting approach; and, ii) it takes, approximately, 20 times more iterations to converge. Both are consequences of the fact that a higher dimensional \textit{constrained} optimization problem is being solved.

\subsection{Incremental initialization approach for the MSA-PEM}

Here, we revisit the setup of the first numerical example presented in \cite{farina_identification_2012}. A dataset with $N=500$ samples is generated using the equation:
\begin{subequations}
\label{eq:example_FarinaPiroddi_y}
\begin{align}
    \bar{y}[k] &= \theta_1 u[k-1] u[k-2] + \theta_2 u[k-1] \bar{y}[k-1] \\
    y[k] &= \bar{y}[k] + v[k].
\end{align}
\end{subequations}
The data generator parameters are $\theta_1=0.6$ and $\theta_2=-0.5$, $\bar{y}$ represents the noiseless output and $v$ the white output noise. The input $u$ is generated by the following AR process:
\begin{equation}
    \label{eq:example_FarinaPiroddi_u}
    u[k] = 0.99 u[k-1] + 0.1 \eta[k],
\end{equation}
where $\eta$ is white Gaussian noise with variance one. The output noise $v$ has variance $0.25 \lambda_y^2$, where $\lambda_y^2=0.18^2$ is the output variance.

A model with the same structure as the data generator is estimated using MSA-PEM. Fig.~\ref{fig:smoothness_ex5} displays the cost function and the result of parameter estimation for several initial optimization conditions. Besides the results of the vanilla implementation, in \textcolor{blue}{{blue}}, we also show, in \textcolor{black}{{black}}, the refinement proposed in~\cite{farina_identification_2012}. For this refinement, the prediction starts from the one-step-ahead solution, and in each iteration increases $K$ by one. That is, the $K$ MSA-PEM problem is initialized from the (partial) solution of the $K-1$ MSA-PEM estimation. The intermediary results for the refined implementation are displayed in the plot. With the algorithm being iterated up to the $K$-th iteration.

The effect of increasing the simulation length can be observed in this example, with a larger $K$ yielding less smooth cost functions. The refinement proposed in~\cite{farina_identification_2012} allows the estimation to approach the desired asymptotic statistical properties while optimizing more complicated cost functions progressively. The results in Fig~\ref{fig:smoothness_ex5}, show  this refinement yield improvements over the vanilla implementation, maybe because it avoids optimizing the harder optimization problems from a distant initial estimate.

\section{Equality-constrained optimization solver}
\label{sec:tr-sqp}

This appendix describes the implementation of the solver described in~\cite{lalee_implementation_1998}, which is used in the numerical examples. The method is able to solve large-scale equality-constrained optimization problems:
\begin{align}
  \min_{\boldsymbol{\theta}} & \,\,V(\boldsymbol{\theta}), \label{eq:eqp}\\
   \text{subject to } & \mathbf{c}(\boldsymbol{\theta}) = 0 \nonumber
\end{align}
where $V$: $\mathbb{R}^n \rightarrow \mathbb{R}$ and
$\mathbf{c}$: $\mathbb{R}^n \rightarrow \mathbb{R}^m$
are twice continuously differentiable functions.

The algorithm solves a sequence of Taylor approximations
in order to gradually converge towards a \textit{local} solution of~(\ref{eq:eqp}).
At the $k$-th iteration, the algorithm builds a local model
around the current iterate $\boldsymbol{\theta}_k$, computes
the update  $\mathbf{p}_k$ by solving a \textit{quadratic programming} problem
and then updates the solution:
\begin{equation}
    \boldsymbol{\theta}_{k+1} = \boldsymbol{\theta}_{k}+ \mathbf{p}_{k}.
\end{equation}
The algorithm is a \textit{trust-region method}~\cite{conn_trustregion_2000}, in the sense that the update must respect $\|\mathbf{p}_{k}\| \leq \Delta_k$, for which $\Delta_k$ is known as \textit{trust radius} and  should reflect the trust the algorithm has on the local approximation of the cost function. If the current local approximation is not a good one, the method does not allow
a very large step to be taken.

\subsection{Quadratic programming subproblem}

At each iteration, in order to compute the step update $\mathbf{p}_k$, the method  solves  the trust-region quadratic programming (QP) sub-problem:
\begin{eqnarray}
  \label{eq:qp_subprob}
  \min_{\mathbf{p}} && \nabla V(\boldsymbol{\theta}_k)^T \mathbf{p} + \frac{1}{2} \mathbf{p}^T \nabla^2_{\theta\theta}
            \mathcal{L}(\boldsymbol{\theta}_k, \boldsymbol{\lambda}_k)^T \mathbf{p}, \\
   \text{subject to: } && J(\boldsymbol{\theta}_k)\mathbf{p} + \mathbf{c}(\boldsymbol{\theta}_k) = \mathbf{r}_k; \nonumber \\
   && \|\mathbf{p}\| \le \Delta_k, \nonumber
\end{eqnarray}
where $\boldsymbol{\theta}_k$ is the current iterate, $\boldsymbol{\lambda}_k$ is the current approximation of Lagrange multipliers, $\nabla V(\boldsymbol{\theta}_k)^T$ is the gradient of $V$  and $J(\boldsymbol{\theta}_k)$ is the Jacobian matrix of $\mathbf{c}$. As in Appendix~\ref{sec:comp-deriv}, $\mathcal{L}(\boldsymbol{\theta}_k, \boldsymbol{\lambda}_k)^T$ denotes the Lagrangian:
\begin{equation}
    \mathcal{L}(\boldsymbol{\theta}_k, \boldsymbol{\lambda}_k) =
  V(\boldsymbol{\theta}_k) +  \boldsymbol{\lambda}_k^T\mathbf{c}(\boldsymbol{\theta}_k),
\end{equation}
and $\nabla^2_{\theta\theta} \mathcal{L}(\boldsymbol{\theta}_k, \boldsymbol{\lambda}_k)^T$
is the Hessian in relation to the variable $\boldsymbol{\theta}$.

The current approximation of the Lagrange multipliers $\boldsymbol{\lambda}_k$ is obtained 
by solving the least-squares problem:
\begin{equation}
  \min_{\boldsymbol{\lambda}} \|\nabla_x\mathcal{L}(\boldsymbol{\theta}_k, \boldsymbol{\lambda}_k)\|^2
  \Rightarrow
  \min_{\boldsymbol{\lambda}} \|\nabla V(\boldsymbol{\theta}_k) + J(\boldsymbol{\theta}_k)\boldsymbol{\lambda}\|^2.
\end{equation}

The vector $\mathbf{r}_k$ is included in the problem in order to guarantee linear and trust-region constraints are always compatible. $\mathbf{r}_k$ is defined as $J(\boldsymbol{\theta}_k)\mathbf{v}_k + \mathbf{c}(\boldsymbol{\theta}_k)$ for a value of $\mathbf{v}_k$ satisfying:
\begin{eqnarray}
  \label{eq:find_tr}
  \min_{\mathbf{v}} && \|J(\boldsymbol{\theta}_k)\mathbf{v} + \mathbf{c}(\boldsymbol{\theta}_k)\|^2, \\\
   \text{subject to } && \|\mathbf{v}\| \le \eta \Delta_k,\nonumber
\end{eqnarray}
for $0<\eta<1$ (in our implementation $\eta=0.8$). For this
choice of $\mathbf{r}_k$ the linear constraints
are always compatible with trust-region constraints.

The sub-steps are solved in rather economical fashion using efficient methods to get an \textit{inexact} solution to each of the sub-problems.  The QP problem~(\ref{eq:qp_subprob}) is solved using a variation of the \textit{dogleg} procedure (described in~\cite{byrd_interior_1999}, p.886) and~(\ref{eq:find_tr}) is solved using the projected conjugate gradient (CG) algorithm~\cite{gould_solution_2001}.

\subsection{Implementation details}

Here we define the \textit{merit function}, $\phi$,  which combines the constraints and the objective function into a single number that can be used to compare two points and to reject or accept a given step. It is given by:
\begin{equation}
\phi(\boldsymbol{\theta}; \mu) = V(\boldsymbol{\theta}) + \mu \|\mathbf{c}(\boldsymbol{\theta})\|,
\end{equation}
where the \textit{penalty parameter} $\mu$ is updated
through the iterations. This parameter must increase monotonically for the algorithm to converge. Some additional guidelines are provided in~\cite{byrd_interior_1999}, p.891.

The selection of the trust radius $\Delta_k$ and the step rejection mechanism
are based on the ratio:
\begin{equation}
\rho_k = \frac{\phi(\boldsymbol{\theta}_k; \mu) -  \phi(\boldsymbol{\theta}_k + \mathbf{p}_k, \mu)}{q_{\mu}(0) -  q_{\mu}(p_k)},
\end{equation}
between the reduction of the merit function and the reduction predicted by the local (Taylor approximation) model:
\begin{multline*}
q_{\mu}(p) = \nabla V(\boldsymbol{\theta}_k)^T p + \frac{1}{2} p^T \nabla^2_{xx} \mathcal{L}(\boldsymbol{\theta}_k, \boldsymbol{\lambda}_k) p \\+ \mu\|c(\boldsymbol{\theta}_k)+J(\boldsymbol{\theta}_k)p\|.
\end{multline*}
The ratio measures the agreement between the observed and expected reduction and is used as decision variable when accepting or rejecting the update and enlarging or reducing the trust radius $\Delta_k$.

\subsection{Algorithm Overview}
The full algorithm is summarized next.

\begin{alg}[Trust-region sequential QP solver]
  At each iteration, until some stop criterion is met (e.g. \newline
  ${\|\nabla_x\mathcal{L}(\boldsymbol{\theta}_k, \boldsymbol{\lambda}_k)\|_{\infty} < 10^{-8}}$),
  repeat:
\begin{enumerate}
\item Compute $V(\boldsymbol{\theta}_k)$, $\nabla V(\boldsymbol{\theta}_k)$, $c(\boldsymbol{\theta}_k)$
  and $J(\boldsymbol{\theta}_k)$;
\item Compute least squares Lagrange multipliers $\boldsymbol{\lambda}_k$;
\item Compute $\nabla^2_{xx} \mathcal{L}(\boldsymbol{\theta}_k, \boldsymbol{\lambda}_k)$;
\item Apply dogleg method in order to compute $v_k$ and $r_k$ (such that the resulting
  problem is feasible);
\item Compute $p_k$ using the projected CG method;
\item Choose penalty parameter $\mu_k$;
\item Compute reduction ratio $\rho_k$;
\item Accept or reject step $p_k$ using $\rho_k$ as decision variable;
\item Enlarge or reduce trust-radius using $\rho_k$ as decision variable.
\end{enumerate}
\end{alg}

\section*{Code avaibility}
The code for reproducing the examples is available in: \href{https://github.com/antonior92/MultipleShootingPEM.jl}{github.com/antonior92/MultipleShootingPEM.jl}.

\end{appendices}

\section*{Acknowledgements}

This work has been supported by the Brazilian agencies \textit{CAPES - Coordenacão de Aperfeiçoamento de Pessoal de N\'ivel Superior} (Finance Code: 001), \textit{CNPq - Conselho Nacional de Desenvolvimento Cient\'ifico e Tecnológico} (contract number: 303412/2019-4, 200931/2018-0 and 142211/2018-4) and \textit{FAPEMIG - Fundação de Amparo \`a Pesquisa do Estado de Minas Gerais} (contract number: TEC 1217/98), by the Swedish Research Council (VR) via the projects \emph{NewLEADS -- New Directions in Learning Dynamical Systems} (contract number: 621-2016-06079) and \emph{Learning flexible models for nonlinear dynamics} (contract number: 2017-03807), and by the Swedish Foundation for Strategic Research (SSF) via the project \emph{ASSEMBLE} (contract number: RIT15-0012).


\begin{thebibliography}{10}
\bibitem{ljung_system_1998}
L.~Ljung, {\em System Identification: Theory for the User}.
\newblock {Prentice Hall}, second~ed., 1999.

\bibitem{ljung_convergence_1978}
L.~Ljung, ``Convergence analysis of parametric identification methods,'' {\em
  IEEE Transactions on Automatic Control}, vol.~23, pp.~770--783, Oct. 1978.

\bibitem{ribeiro_parallel_2018}
A.~H. Ribeiro and L.~A. Aguirre, ``''{{Parallel Training Considered
  Harmful}}?'': {{Comparing}} series-parallel and parallel feedforward network
  training,'' {\em Neurocomputing}, vol.~316, pp.~222--231, Nov. 2018.

\bibitem{aguirre_prediction_2010}
L.~A. Aguirre, B.~H. Barbosa, and A.~P. Braga, ``Prediction and simulation
  errors in parameter estimation for nonlinear systems,'' {\em Mechanical
  Systems and Signal Processing}, vol.~24, no.~8, pp.~2855--2867, 2010.

\bibitem{su_longterm_1992}
H.~T. Su, T.~J. McAvoy, and P.~Werbos, ``Long-term predictions of chemical
  processes using recurrent neural networks: {{A}} parallel training
  approach,'' {\em Industrial \& Engineering Chemistry Research}, vol.~31,
  no.~5, pp.~1338--1352, 1992.

\bibitem{piroddi_simulation_2008}
L.~Piroddi, ``Simulation {{Error Minimisation Methods}} for {{NARX Model
  Identification}},'' {\em International Journal of Modelling, Identification
  and Control}, vol.~3, no.~4, pp.~392--403, 2008.

\bibitem{paduart_identification_2010}
J.~Paduart, L.~Lauwers, J.~Swevers, K.~Smolders, J.~Schoukens, and R.~Pintelon,
  ``Identification of nonlinear systems using {{Polynomial Nonlinear State
  Space}} models,'' {\em Automatica}, vol.~46, pp.~647--656, Apr. 2010.

\bibitem{schoukens_identification_2017}
M.~Schoukens and K.~Tiels, ``Identification of block-oriented nonlinear systems
  starting from linear approximations: {{A}} survey,'' {\em Automatica},
  vol.~85, pp.~272--292, Nov. 2017.

\bibitem{piroddi_identification_2003}
L.~Piroddi and W.~Spinelli, ``An identification algorithm for polynomial
  {{NARX}} models based on simulation error minimization,'' {\em International
  Journal of Control}, vol.~76, pp.~1767--1781, Nov. 2003.

\bibitem{eckhard_cost_2017}
D.~Eckhard, A.~S. Bazanella, C.~R. Rojas, and H.~Hjalmarsson, ``Cost function
  shaping of the output error criterion,'' {\em Automatica}, vol.~76,
  pp.~53--60, Feb. 2017.

\bibitem{bock_recent_1983}
H.~Bock, ``Recent {{Advances}} in {{Parameter Identification Problems}} for
  {{ODE}},'' {\em Numerical Treatment of Inverse Problems in Differential and
  Integral Equations}, pp.~95--121, 1983.

\bibitem{baake_fitting_1992}
E.~Baake, M.~Baake, H.~Bock, and K.~Briggs, ``Fitting ordinary differential
  equations to chaotic data,'' {\em Physical Review A}, vol.~45, no.~8,
  p.~5524, 1992.

\bibitem{sarode_embedded_2015}
K.~D. Sarode, V.~R. Kumar, and B.~Kulkarni, ``Embedded {{Multiple Shooting
  Methodology}} in a {{Genetic Algorithm Framework}} for {{Parameter
  Estimation}} and {{State Identification}} of {{Complex Systems}},'' {\em
  Chemical Engineering Science}, vol.~134, pp.~605--618, 2015.

\bibitem{bock_multiple_1984}
H.~G. Bock and K.-J. Plitt, ``A multiple shooting algorithm for direct solution
  of optimal control problems,'' {\em IFAC Proceedings Volumes}, vol.~17,
  no.~2, pp.~1603--1608, 1984.

\bibitem{carraro_indirect_2014}
T.~Carraro, M.~Geiger, and R.~Rannacher, ``Indirect {{Multiple Shooting}} for
  {{Nonlinear Parabolic Optimal Control Problems}} with {{Control
  Constraints}},'' {\em SIAM Journal on Scientific Computing}, vol.~36, no.~2,
  pp.~A452--A481, 2014.

\bibitem{geisert_trajectory_2016}
M.~Geisert and N.~Mansard, ``Trajectory {{Generation}} for {{Quadrotor Based
  Systems Using Numerical Optimal Control}},'' in {\em 2016 {{IEEE}}
  International Conference on Robotics and Automation ({{ICRA}})},
  pp.~2958--2964, {IEEE}, 2016.

\bibitem{vanmulders_two_2010}
A.~Van~Mulders, J.~Schoukens, M.~Volckaert, and M.~Diehl, ``Two nonlinear
  optimization methods for black box identification compared,'' {\em
  Automatica}, vol.~46, pp.~1675--1681, Oct. 2010.

\bibitem{ribeiro_shooting_2017}
A.~H. Ribeiro and L.~A. Aguirre, ``Shooting {{Methods}} for {{Parameter
  Estimation}} of {{Output Error Models}},'' {\em IFAC-PapersOnLine}, vol.~50,
  pp.~13998--14003, July 2017.

\bibitem{farina_simulation_2011}
M.~Farina and L.~Piroddi, ``Simulation error minimization identification based
  on multi-stage prediction,'' {\em International Journal of Adaptive Control
  and Signal Processing}, vol.~25, no.~5, pp.~389--406, 2011.

\bibitem{terzi_learning_2018}
E.~Terzi, L.~Fagiano, M.~Farina, and R.~Scattolini, ``Learning multi-step
  prediction models for receding horizon control,'' in {\em 2018 {{European
  Control Conference}} ({{ECC}})}, pp.~1335--1340, June 2018.

\bibitem{farina_identification_2012}
M.~Farina and L.~Piroddi, ``Identification of polynomial input/output recursive
  models with simulation error minimisation methods,'' {\em International
  Journal of Systems Science}, vol.~43, no.~2, pp.~319--333, 2012.

\bibitem{noel_greybox_2018}
J.\,P.\,No{\"e}l and J.~Schoukens, ``Grey-box state-space identification of
  nonlinear mechanical vibrations,'' {\em International Journal of Control},
  vol.~91, pp.~1118--1139, May 2018.

\bibitem{boyd_fading_1985}
S.~Boyd and L.~Chua, ``Fading memory and the problem of approximating nonlinear
  operators with {{Volterra}} series,'' {\em IEEE Transactions on Circuits and
  Systems}, vol.~32, pp.~1150--1161, Nov. 1985.

\bibitem{nesterov_introductory_1998}
Y.~Nesterov, {\em Introductory {{Lectures On Convex Programming}}}.
\newblock {Springer Science \& Business Media}, 1998.

\bibitem{pascanu_difficulty_2013}
R.~Pascanu, T.~Mikolov, and Y.~Bengio, ``On the {{Difficulty}} of {{Training
  Recurrent Neural Networks}},'' in {\em Proceedings of the 30th
  {{International Conference}} on {{International Conference}} on {{Machine
  Learning}}}, vol.~28, pp.~1310--1318, 2013.

\bibitem{ribeiro_exploding_2020}
A.~H. Ribeiro, K.~Tiels, L.~A. Aguirre, and T.~B. Sch{\"o}n, ``Beyond exploding
  and vanishing gradients: Attractors and smoothness in the analysis of
  recurrent neural network training,'' in {\em Proceedings of the 23rd
  {{International Conference}} on {{Artificial Intelligence}} and
  {{Statistics}} ({{AISTATS}})}, vol.~108, pp.~2370--2380, 2020.

\bibitem{rudin_principles_1964}
W.~Rudin, {\em Principles of Mathematical Analysis}.
\newblock International Series in Pure and Applied Mathematics, {McGraw-Hill},
  1964.

\bibitem{lalee_implementation_1998}
M.~Lalee, J.~Nocedal, and T.~Plantenga, ``On the implementation of an algorithm
  for large-scale equality constrained optimization,'' {\em SIAM Journal on
  Optimization}, vol.~8, no.~3, pp.~682--706, 1998.

\bibitem{virtanen_scipy_2020}
P.~Virtanen, R.~Gommers, T.~E. Oliphant, M.~Haberland, T.~Reddy, D.~Cournapeau,
  E.~Burovski, P.~Peterson, W.~Weckesser, J.~Bright, S.~J. {van der Walt},
  M.~Brett, J.~Wilson, K.~J. Millman, N.~Mayorov, A.~R.~J. Nelson, E.~Jones,
  R.~Kern, E.~Larson, C.~J. Carey, {\.I}.~Polat, Y.~Feng, E.~W. Moore,
  J.~VanderPlas, D.~Laxalde, J.~Perktold, R.~Cimrman, I.~Henriksen, E.~A.
  Quintero, C.~R. Harris, A.~M. Archibald, A.~H. Ribeiro, F.~Pedregosa, P.~{van
  Mulbregt}, and S.~. Contributors, ``{{SciPy}} 1.0--{{Fundamental Algorithms}}
  for {{Scientific Computing}} in {{Python}},'' {\em Nature Methods}, 2020.

\bibitem{may_simple_1976}
R.~M. May, ``Simple mathematical models with very complicated dynamics,'' {\em
  Nature}, vol.~261, no.~5560, pp.~459--467, 1976.
  
\bibitem{newey_large_1994a}
W.~K. Newey and D.~McFadden, ``Large sample estimation and hypothesis
  testing,'' {\em Handbook of econometrics}, vol.~4, pp.~2111--2245, 1994.

\bibitem{ljung_asymptotic_1980}
L.~Ljung and P.~E. Caines, ``Asymptotic normality of prediction error
  estimators for approximate system models,'' {\em Stochastics}, vol.~3,
  pp.~29--46, Jan. 1980.

\bibitem{conn_trustregion_2000}
A.~R. Conn, N.~I.~M. Gould, and P.~L. Toint, {\em Trust-Region Methods}.
\newblock {{MPS}}-{{SIAM}} Series on Optimization, {Philadelphia, PA}: {Society
  for Industrial and Applied Mathematics}, 2000.

\bibitem{byrd_interior_1999}
R.~H. Byrd, M.~E. Hribar, and J.~Nocedal, ``An interior point algorithm for
  large-scale nonlinear programming,'' {\em SIAM Journal on Optimization},
  vol.~9, no.~4, pp.~877--900, 1999.

\bibitem{gould_solution_2001}
N.~I. Gould, M.~E. Hribar, and J.~Nocedal, ``On the solution of equality
  constrained quadratic programming problems arising in optimization,'' {\em
  SIAM Journal on Scientific Computing}, vol.~23, no.~4, pp.~1376--1395, 2001.

\bibitem{lecun_efficient_1998}
Y.~LeCun, L.~Bottou, G.~B. Orr, and K.-R. M{\"u}ller, ``Efficient
  {{BackProp}},'' in {\em Neural {{Networks}}: {{Tricks}} of the {{Trade}}},
  Lecture {{Notes}} in {{Computer Science}}, pp.~9--50, {Springer, Berlin,
  Heidelberg}, 1998.

\bibitem{dvoretzky_asymptotic_1956}
A.~Dvoretzky, J.~Kiefer, and J.~Wolfowitz, ``Asymptotic {{Minimax Character}}
  of the {{Sample Distribution Function}} and of the {{Classical Multinomial
  Estimator}},'' {\em The Annals of Mathematical Statistics}, vol.~27,
  pp.~642--669, Sept. 1956.
\end{thebibliography}
\end{document}